%% file: sample-manuscript.tex
\documentclass[manuscript]{acmart}
\AtBeginDocument{%
  }

\acmJournal{TOSEM}
\setcopyright{acmlicensed}
\copyrightyear{2026}
\acmYear{2026}
\acmDOI{XXXXXXX.XXXXXXX}
\acmConference[Conference acronym 'XX]{Make sure to enter the correct
  conference title from your rights confirmation email}{June 03--05,
  2018}{Woodstock, NY}
\acmISBN{978-1-4503-XXXX-X/2026/06}

\usepackage{listings}
\lstdefinelanguage{json}{
  basicstyle=\ttfamily\scriptsize,
  string=[s]{"}{"},
  stringstyle=\color{blue},
  comment=[l]{//},
  morestring=[b]',
  literate=
    *{0}{{{\color{gray}0}}}{1}
     {1}{{{\color{gray}1}}}{1}
     {2}{{{\color{gray}2}}}{1}
     {3}{{{\color{gray}3}}}{1}
     {4}{{{\color{gray}4}}}{1}
     {:}{{{\color{black}:}}}{1}
     {,}{{{\color{black},}}}{1}
}

\usepackage{multirow}
\usepackage{graphicx}

\usepackage{booktabs, colortbl, xcolor}

\definecolor{trainbg}{HTML}{F2DCDB}
\definecolor{testbg}{HTML}{D8E4BC}
\begin{document}

\title{Trajectory-Aware Benchmark Subset Selection for Cost-Efficient Software Engineering Agent Regression Testing}

\author{Mahmoud Ayyad}
\email{mahmoud.ayyad@queensu.ca}
\orcid{0009-0006-0365-3180}
\affiliation{%
  \institution{Queen's University}
  \country{Canada}
}

\author{Zehao Wang}
\email{w\_zeha@encs.concordia.ca}
\orcid{0000-0001-6053-6278}
\affiliation{%
  \institution{Concordia University}
  \country{Canada}
}

\author{Jiho Shin}
\email{jiho.shin@queensu.ca}
\orcid{0000-0001-8829-3773}
\affiliation{%
  \institution{Queen's University}
  \country{Canada}
}

\author{Ying Zou}\
\email{ying.zou@queensu.ca}
\orcid{0000-0002-5335-0261}
\affiliation{%
  \institution{Queen's University}
  \country{Canada}
}

\author{Bram Adams}
\email{bram.adams@queensu.ca}
\orcid{0000-0001-7213-4006}
\affiliation{%
  \institution{Queen's University}
  \country{Canada}
}


\begin{abstract}

\input{Sections/Abstract}
\end{abstract}




\maketitle

\input{Sections/Introduction}

\input{Sections/RelatedWork}

\input{Sections/Methodology}

\input{Sections/Setup}
\input{Sections/Results}
\input{Sections/Implications}

\input{Sections/ThreatsToValidity}

\input{Sections/Conclusion}


\begin{acks}
We thank the Nebius team for publicly releasing the SWE-rebench-OpenHands-Trajectories dataset used in our study.
\end{acks}

\bibliographystyle{ACM-Reference-Format}
\bibliography{Bib}

\end{document}

%% file: Sections/Abstract.tex
Autonomous software engineering agents (SWE-agents) automate complex coding tasks such as fixing issues and implementing new features.
While evaluating SWE-agents requires running large benchmarks over many real-world software tasks, each agent update may require re-running the full benchmark again to detect regressions and improvements, at a cost of hundreds of millions of LLM tokens per run, which makes evaluation a bottleneck during development.
One solution is to evaluate only a subset of benchmark instances. Yet, simple approaches, such as random sampling or stratified random sampling based on past pass/fail outcomes, risk producing high variance and unrepresentative subsets.
To address this issue, we turn to agent trajectories, the step-by-step record of the actions an agent took and the observations it made while working on an instance. Prior work analyzes trajectories to characterize agent behavior and failure modes, but, to our knowledge, none uses them to select which instances to evaluate. We propose a trajectory-aware subset selection approach that replaces random sampling with deterministic selection based on trajectory embeddings.
We first group test set instances by their test outcome in a recent full test run to preserve the historical pass/fail rate, then select the subset using the trajectory's embedding space.

We systematically evaluate 76 subset selection configurations, including random sampling, embedding-based selection, clustering-based selection, and hybrid shortlist-then-subsample strategies, across three regression scenarios: same-configuration reruns, model and configuration changes, and agent framework changes. Our best trajectory-aware method is the one selecting benchmark instances closest to the centroid of each outcome group in the embedding space.
It achieves the lowest estimation error among all methods we evaluate, where the error is the gap between the resolve rate measured on the subset and the resolve rate of the full benchmark. For instance, when evaluating a given agent version on a selected subset of only 5\% or 10\% of the test instances, our approach reduces the average estimation error by 3--11\% and the worst-case error, the largest difference between the subset's resolve rate and the full benchmark's resolve rate on any single run, by 4--11\% relative to the typical draw and 38--46\% relative to the 95th-percentile draw of the strongest baseline.
Our results show that a 10\% trajectory-aware subset keeps the median estimation error below 5\% while cutting token cost by roughly 90\%.

%% file: Sections/Introduction.tex
\section{Introduction}
\label{sec:introduction}

Autonomous software engineering agents (SWE-agents) are increasingly capable of repairing programs~\cite{zhang2024autocoderover}, debugging software systems~\cite{zhong2024debuglikehumanlarge}, and generating and executing tests~\cite{huang2024agentcoder}.
Benchmarks such as SWE-Bench~\cite{swbench} and its curated subset, SWE-Bench Verified~\cite{swbenchverified}, have been widely used for testing SWE-agents, with newer, harder benchmarks such as SWE-Bench Pro~\cite{deng2025swebenchproaiagents} continuing this line. Each benchmark consists of a set of instances, where each instance typically is a real GitHub issue in an existing codebase. The agent attempts to resolve the GitHub issue by taking a series of actions, such as reading files and modifying code, generating a trajectory. To resolve each instance, the agent must output a final code fix that makes the instance's previously failing tests pass while keeping the previously passing tests green.

Major SWE-agent frameworks, such as OpenHands~\cite{2025openhands}, have been using SWE-Bench instances for quality monitoring during development.
Running such benchmarks is costly, both in terms of time and money.
This is because each evaluation requires executing hundreds of instances (e.g., 500 in SWE-Bench Verified), and each instance involves a full agent loop with multiple LLM API calls, tool invocations, and code-test execution steps.
As a result, repeatedly evaluating every SWE-agent update, for instance in a CI workflow or even ad hoc on the framework developers' end, becomes impractical. Yet, developers continuously modify models, prompts, and agent architectures, and each change may require regression testing to detect improvements or performance drops~\cite{fan2025sweeffi}, which calls for a cheaper check than a full benchmark run.

One mitigation for this evaluation cost is therefore to evaluate on a smaller subset of the benchmark.
Prior work has explored curated subsets based on instance difficulty and solvability~\cite{swbenchverified, ndzomga2026efficientbenchmarking}.
For example, the developers behind OpenHands have recognized the necessity of smaller subsets~\cite{2025openhands}. Their evaluation infrastructure supports limiting runs to smaller instance counts (e.g., 50 or 200 out of 500 in SWE-bench Verified), yet the subset is selected via a fixed-seed random sample.\footnote{See \texttt{prepare\_dataset} at line 70 of \url{https://github.com/OpenHands/benchmarks/blob/4e5469e0caaf54d1ad827d18b524bdfb79d58430/benchmarks/utils/dataset.py\#L70C9-L70C80}, which calls \texttt{dataset.sample(n=n\_limit, random\_state=42)}.\label{fn:openhands}}
While traditional software engineering has long used historical pass/fail outcomes as an effective signal for the test-suite reduction and prioritization~\cite{kim2002history, elbaum2014}, to the best of our knowledge, no technique thus far has exploited this idea in the context of SWE-agent regression testing.

Motivated by the prior work, one direction is to construct subsets that preserve the full test suite's overall difficulty using agents' historical pass/fail outcomes, even though such subset selection is challenging~\cite{polo2024tinybenchmarksevaluatingllmsfewer}. In particular, subset selection based only on static instance properties or final pass/fail labels ignores what happens during execution: instances with the same outcome can reflect very different agent behaviors and failure modes.
Recent studies of agent trajectories show that similar failures can happen for different reasons, such as repetitive command loops or failure to run the test suite~\cite{bouzenia2025, majgaonkar2025}. In other words, building an evaluation subset that accurately reflects the full test suite requires behavioral information extracted directly from the agent's step-by-step trajectories on prior runs. Furthermore, subsets built without trajectory data can produce unstable estimates. At small subset sizes, random sampling has high variance and may over-represent rare edge cases rather than typical execution patterns~\cite{polo2024tinybenchmarksevaluatingllmsfewer}. Stratifying by outcome reduces this variance but does not remove it, since stratification controls the variation along the variable used to form the groups~\cite{neyman1934representative}: within each outcome group, instances are still drawn at random. Our approach keeps this stratification but replaces the random draw.

Hence, we propose a \textit{trajectory-aware subset selection} approach that combines historical execution outcomes with the actions the agent took while solving each instance (its trajectory), for example, the number of steps it took, the amount of tests ran, and the number of produced errors. Our approach selects a subset of benchmark instances whose execution profile matches the full test suite, so the subset reflects the same workload the agent faces on the full run. We first group benchmark instances (i.e., individual issues) by the historical pass/fail outcomes of the agent under development (its earlier versions or configurations), a step we call outcome grouping, to preserve the full test suite's overall difficulty distribution. We then apply a sanitization pipeline to remove \textit{outcome leakage}: words in the trajectory
that reveal whether the run passed or failed (e.g., 
``tests passed'', ``exit code 0''), along with
repository-specific terms that identify the instance rather than describe the agent's behavior. Finally, we transform the sanitized trajectories into semantic embeddings, which allow us to measure the distance between different agent behaviors and apply geometric selection algorithms. Inspired by traditional test prioritization approaches based on execution traces~\cite{jabbar2022test2vec, AlSharif2024AbstractTraceTU, kim2002history}, we take the trajectories from the most recent full runs and use them to select a small set of instances that represent the full benchmark.

To validate our approach under realistic development conditions, where developers iterate on agents by changing models, configurations, or frameworks and rerun the benchmark to check for regressions, we empirically analyze subset selection performance on 31,779 publicly available trajectories from 58 runs spanning five agent frameworks. The data covers three distinct regression testing scenarios: (1) 45 reruns of the same OpenHands agent on SWE-Rebench (3,188 instances, 25,279 trajectories), (2) 6 runs of OpenHands with different models/configurations on SWE-Bench Verified (3,000 trajectories), and (3) 7 runs across five distinct SWE-agents on SWE-Bench Verified (3,500 trajectories). Using this data, we investigate four research questions:

\vspace{15pt}

\noindent \textbf{RQ1: How effective are baseline subset selection approaches at approximating full-test suite resolve rates?}
To establish a baseline, we evaluate several simple selection methods. We test methods such as ``uniform random sampling" (picking instances at random without prior knowledge), ``repository stratification" (grouping instances by their source repository), and ``outcome-based stratification" (grouping instances by their historical outcomes, such as pass, fail, or flaky, and sampling proportionally from those groups). While grouping by past test outcomes reduces the median error by 31--51\% relative to random selection, all simple methods prove unreliable: at subset sizes of 5\% and 10\%, even the strongest baseline has a worst-case error of up to 14\% on a typical draw and up to 35\% on its worst draw.

\noindent \textbf{RQ2: To what extent do trajectory-aware embedding methods improve subset selection compared to the strongest baseline?}
We investigate whether selecting instances based on the agent's step-by-step behavior (trajectory) provides a more reliable subset. After grouping the instances by their past test outcomes (outcome grouping), we transform the agents' trajectories from recent runs into embeddings. Instead of pulling random instances, we apply geometric selection by choosing the instances closest to the center of each group. At subset sizes of 5\% and 10\%, the method reduces the average estimation error by 3--11\% while cutting the worst-case error by 4--11\% relative to the typical draw and 38--46\% relative to the 95th-percentile draw of the strongest baseline approach. Using a 10\% subset, our approach reduces test suite execution by 90\% while maintaining median estimation error below 5\%.

\noindent \textbf{RQ3: What drives the improvement of trajectory-aware methods over the baselines?}
We measure the impact of outcome grouping and
geometric selection on the median estimation error and worst-case error reduction reported in RQ2, both individually and combined. We find that both components are necessary. Removing outcome grouping increases the average estimation error by 30--37\%, while a fixed selection rule that uses past outcomes but no embeddings is not consistently better than random sampling and falls behind it as the subset grows. Replacing exact geometric selection with random sampling from a candidate pool worsens the average estimation error by 21--26\%. Only the combination of outcome grouping and geometric selection improves the average accuracy by 3--11\% and reduces worst-case error by 4--11\% relative to the typical draw and 38--46\% relative to the 95th-percentile draw of the strongest baseline approach, as it preserves the balance between hard and easy instances, and geometric selection over the embedding space captures representative behavior without random variance.

\noindent \textbf{RQ4: How much does subset selection
reduce evaluation cost?}
Whether selecting a 10\% subset saves 90\% of the benchmark execution cost depends
on which instances it contains, since some instances
cost more to evaluate than others. We measure the
total token cost of 3{,}000 trajectories from six runs
of the same agent (i.e., OpenHands) under different models and configurations, then draw 10{,}000 random $k$\% subsets per run to estimate the actual token cost savings. We find that our approach is cost-neutral: it favors neither cheap nor expensive instances, so a k\% subset consumes
k\% of the full token cost on average. The accuracy and worst-case error gains from RQ2 therefore come with proportional cost savings. A 10\% subset reduces token consumption from 3.44B to 345M, a 90\% saving, while keeping median estimation error below 5\%.

The contributions of this paper are as follows:
\begin{itemize}
    \item We propose the first trajectory-aware subset selection approach for SWE-agent regression testing that combines deterministic geometric selection with historical outcome grouping over sanitized trajectory embeddings.
    \item We develop a four-phase sanitization pipeline that removes outcome leakage from agent trajectories while preserving behaviorally meaningful signals. We additionally provide a unified parsing schema that standardizes heterogeneous trajectory formats across five agent frameworks into a common representation.
    \item We evaluate 76 subset selection configurations for SWE-agent regression testing across three regression scenarios (same-configuration reruns, model and configuration changes, and agent framework changes) with strict temporal cross-validation.
    \item We release the complete dataset of standardized trajectories, the unified parsing schema, and selection algorithms to facilitate replication\footnote{\url{www.github.com/SAILResearch/swe-agent-subset-selection}}.
\end{itemize}

The remainder of this paper is organized as follows.
Section~\ref{sec:background} introduces background concepts.
Section~\ref{sec:relatedwork} introduces related works. 
Section~\ref{sec:methodology} details our approach. 
Section~\ref{sec:eval_setup} outlines the experimental setup. 
Section~\ref{sec:results} presents the findings. 
Section~\ref{sec:discussion} discusses the implications and practical insights from our findings. 
Section~\ref{sec:threatstovalidity} examines validity threats, 
and Section~\ref{sec:conclusion} concludes.

%% file: Sections/RelatedWork.tex
\section{Background}
\label{sec:background}

\subsection{SWE-Agents}
\label{sec:background-swe-agents}
An AI agent is an autonomous software system driven by an AI model that can reason, use tools, and run in a loop to achieve a specific goal. In each iteration of the loop, the agent chooses an action, executes it through a tool (e.g., a shell or a file editor), and observes the result, which informs its next action. In the context of software engineering agents (SWE-agents), the goal is to work on software. Given an issue report and access to the corresponding codebase, a SWE-agent reads files, modifies code, and runs commands and tests until it produces a candidate code fix. There are several SWE-agent frameworks. For instance, AutoCodeRover~\cite{zhang2024autocoderover} iteratively searches the codebase to locate and patch the code responsible for an issue, while OpenHands~\cite{2025openhands} provides an open platform where agents act through a shell, a file editor, and a browser.

\subsection{Agent Trajectories}
\label{sec:background-trajectories}
During execution, SWE-agent frameworks log the agent's sequence of thoughts, actions, and observations, called a \textit{trajectory}. Figure~\ref{fig:trajectory-example} shows the structure of a trajectory: each step in a trajectory records the action the agent took (e.g., a file edit or a shell command) and the observation it received back (e.g., the edit confirmation or the command output).

\begin{figure}[t]
\centering
\begin{minipage}{0.6\linewidth}
\begin{lstlisting}[
  basicstyle=\ttfamily\scriptsize,
  frame=single,
  numbers=none,
  breaklines=true,
  linewidth=\linewidth,
  aboveskip=4pt
]

Step 1
  Thought:     "The issue mentions a bug in
                misc.py, let me inspect it."
  Action:      open codebase/utils/misc.py
  Observation: <file contents>

Step 2
  Thought:     "The comparison operator is
                wrong, I will fix it."
  Action:      edit codebase/utils/misc.py
  Observation: "The file has been edited..."

Step 3
  Thought:     "Let me run the tests to
                verify the fix."
  Action:      run pytest codebase/utils/
  Observation: <test output>

...
Final output: candidate code fix (patch)
\end{lstlisting}
\end{minipage}
\caption{Example of a SWE-agent trajectory. Each step records the agent's thought, the action it took, and the observation it received.}
\label{fig:trajectory-example}
\end{figure}

\section{Related Work}
\label{sec:relatedwork}

\subsection{Behavioral Analysis of Agent Trajectories}
\label{sec:rw-background-trajectories}
A growing line of work analyzes these trajectories to understand agent behavior. Bouzenia and Pradel~\cite{bouzenia2025} study thought-action-result patterns in SWE-agents to characterize decision-making dynamics and failure modes. Ceka et al.~\cite{ceka2026understandingautomatedprogramrepair} trace the full decision-making pipelines of five agents, derive a taxonomy of their decision pathways, and identify bug localization and reproduction-test generation as core behavioral modules. Majgaonkar et al.~\cite{majgaonkar2025} compare successful and failed trajectories, showing that longer and more variable trajectories are often associated with failure. Beyond characterization, recent work scores trajectory quality: AgentLens~\cite{sahoo2026agentlensrevealingluckypass} computes deterministic process scores for SWE-agent trajectories to expose solutions that pass tests despite flawed processes. However, prior trajectory analysis targets interpretability and post-hoc characterization. We instead use trajectories as a selection signal, embedding them to pick a behaviorally representative subset of benchmark instances.

\subsection{Efficient Evaluation of Autonomous SWE-Agents}
SWE-agents are evaluated on benchmarks such as SWE-Bench~\cite{swbench} and the newer, harder SWE-Bench Pro~\cite{deng2025swebenchproaiagents}. Each benchmark consists of a set of instances, where each instance typically is a real GitHub issue in an existing codebase. The fraction of resolved instances is the agent's \textit{resolve rate}. Evaluating an agent means running it on every instance of the benchmark, and this high cost has motivated two lines of research: reducing the resources required to execute a single benchmark instance and reducing the total number of instances to evaluate.

To reduce the cost of executing individual benchmark instances, prior work has focused on lowering token usage and execution time. For example, Fan et al.~\cite{fan2025sweeffi} propose SWE-Effi, a resource-aware evaluation framework that measures agent effectiveness under resource constraints and penalizes excessive token and time consumption. AgentDiet~\cite{agentdiet} reduces inference-time cost by compressing agent trajectories and removing redundant or expired context from the growing interaction history. While these methods reduce the cost per evaluated instance, they do not reduce the number of instances that must be evaluated.

To reduce the total number of instances that require evaluation, researchers have explored evaluating smaller benchmarks. SWE-Bench Lite~\cite{swbench} and the human-validated SWE-Bench Verified~\cite{swbenchverified}, both curated subsets of SWE-Bench, cut cost this way, but they are fixed sets, not methods that adapt the subset per run, and even SWE-Bench Verified remains costly: each evaluation executes 500 instances, each involving a full agent loop with multiple LLM calls and test executions~\cite{fan2025sweeffi}. Closer to our setting, agent frameworks already subsample their benchmarks to cut evaluation cost during development. The OpenHands evaluation harness, for example, limits a run to a fixed-seed random sample of SWE-Bench Verified (e.g. 50 or 200 of 500) with no selection criteria beyond the fixed random seed.\footref{fn:openhands} This is the practice our work targets: our trajectory-aware selection replaces such blind random sampling.

Beyond software engineering agents, prior work on general LLM evaluation has also studied algorithmic benchmark reduction. Zhao et al.~\cite{zhao2024bento} propose BenTo, which embeds instance prompts and selects a subset that maximizes in-context transferability. Bean et al.~\cite{2025scales} introduce Scales++, which shifts attention from question text to the underlying skills each instance tests. Their method uses an LLM to rate the cognitive demands of each instance from its description text alone, embeds these demand ratings, and selects a subset that maximizes diversity in the demand space.

Our work differs from these approaches in both the selection signal and the evaluation objective. BenTo and Scales++ rely on static task properties and target cross-model ranking preservation: they pick instances so that the subset ranks a set of models in the same order as the full benchmark, on multi-task benchmarks with categorical task structure such as MMLU. We instead track a single agent's resolve rate across successive runs of the same benchmark, where each instance is an individual issue report without a predefined category structure, and this rate depends on how the agent behaves rather than on fixed instance properties. Direct application of BenTo and Scales++ to our setting is impractical due to their computational requirements: BenTo requires $\mathcal{O}(n^2)$ LLM calls to build its transferability matrix, exceeding the cost of running the full benchmark, and Scales++, although needing only a one-time LLM pass per instance, still optimizes for ranking a set of models rather than estimating one agent's resolve rate.

\subsection{Execution-Based Test Case Prioritization}
Dynamic selection based on execution data is an established technique in traditional software engineering for improving software testing efficiency, with a large literature surveyed in~\cite{yooAndHarmanRegressionTestingMinimizationSurvey, LOU20191}. Kim and Porter~\cite{kim2002history} showed that historical execution data can support cost-effective test prioritization for regression testing in resource-constrained continuous integration environments. Elbaum et al.~\cite{elbaum2014} extended history-based prioritization to industrial-scale continuous integration at Google, using failure history and execution windows to select and prioritize test suites. Reinforcement learning approaches, such as RETECS~\cite{2017RETECS} and the work of Bagherzadeh and Briand~\cite{bagherzadeh2021}, further learn prioritization policies from sequences of continuous integration outcomes. More recent learning-based approaches scale ML-based prioritization to CI environments using comprehensive feature sets~\cite{Yaraghi_2023}, or rank test cases by semantic similarity to code changes using pretrained language models~\cite{Semanticaware_2024}.

Test2Vec~\cite{jabbar2022test2vec} embeds execution traces into a latent space for test case prioritization and demonstrates that trace-based representations capture behavioral semantics that are missed by code coverage metrics. AbstractTrace~\cite{AlSharif2024AbstractTraceTU} similarly uses execution traces to cluster and prioritize test suites. We adapt the trace-embedding approach to the agent setting by treating agent trajectories as behavioral abstractions for subset selection rather than test prioritization.

%% file: Sections/Methodology.tex
\section{Test Selection Approach}
\label{sec:methodology}

\begin{figure*}[!t]
\centering
\includegraphics[width=\linewidth]{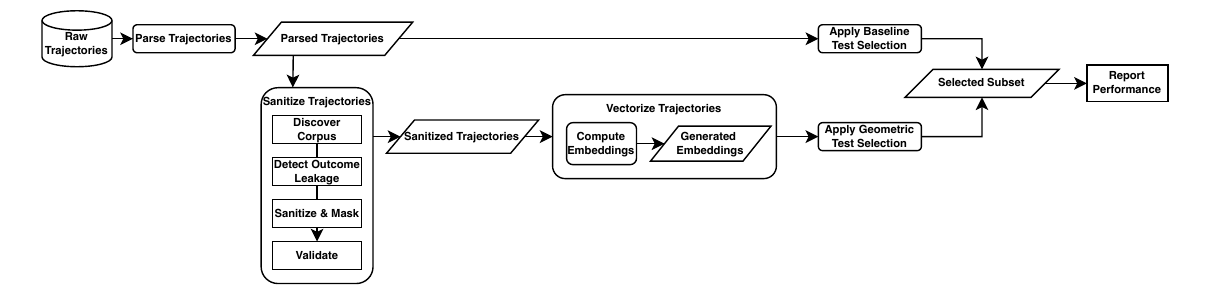}
\captionsetup{labelfont={normalsize}}
\caption{\textbf{Overview of the test selection approach.}}
\label{fig:approach_overview}
\end{figure*}

This section details the end-to-end pipeline for our SWE-agent regression test selection approach, as outlined in Figure \ref{fig:approach_overview}. We first parse raw trajectories into a standardized format as different agent frameworks record trajectories in incompatible ways. Next, we sanitize the parsed trajectories to remove explicit outcome signals, since leaked success or failure tokens would make embeddings cluster by result rather than behavior. We then pass the sanitized trajectories into an embedding model to generate numerical representations. Based on the embeddings of the most recent full regression run, we select a small subset of instances. On later runs, we evaluate only the selected subset and use its resolve rate to estimate the full suite's resolve rate.

\subsection{Parse Trajectories}

\begin{figure*}[t]
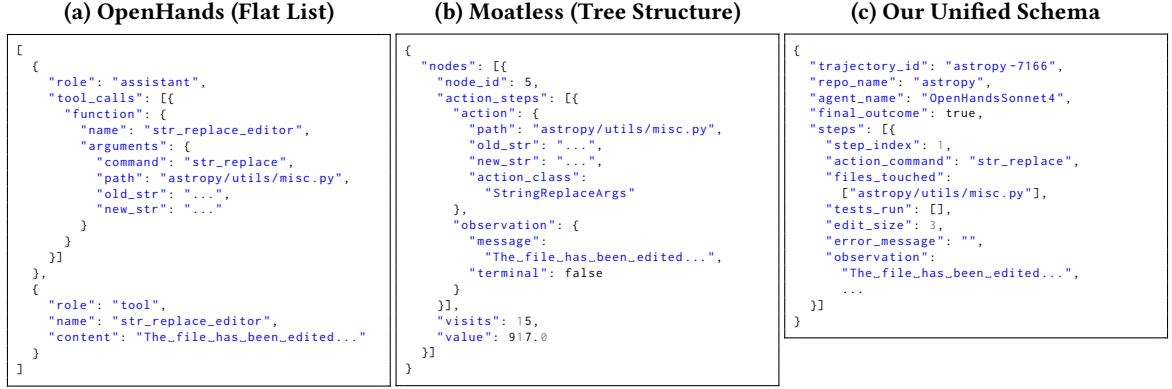

\centering
\begin{minipage}[t]{0.32\textwidth}
\centering
\textbf{(a) OpenHands (Flat List)}
\begin{lstlisting}[
  language=json,
  basicstyle=\ttfamily\tiny,
  frame=single,
  numbers=none,
  breaklines=true,
  breakatwhitespace=true,
  linewidth=\linewidth,
  showlines=true,
  aboveskip=4pt
]
[
  {
    "role": "assistant",
    "tool_calls": [{
      "function": {
        "name": "str_replace_editor",
        "arguments": {
          "command": "str_replace",
          "path": "astropy/utils/misc.py",
          "old_str": "...",
          "new_str": "..."
        }
      }
    }]
  },
  {
    "role": "tool",
    "name": "str_replace_editor",
    "content": "The file has been edited..."
  }
]
\end{lstlisting}
\end{minipage}%
\hfill
\begin{minipage}[t]{0.32\textwidth}
\centering
\textbf{(b) Moatless (Tree Structure)}
\begin{lstlisting}[
  language=json,
  basicstyle=\ttfamily\tiny,
  frame=single,
  numbers=none,
  breaklines=true,
  breakatwhitespace=true,
  linewidth=\linewidth,
  showlines=true,
  aboveskip=4pt
]
{
  "nodes": [{
    "node_id": 5,
    "action_steps": [{
      "action": {
        "path": "astropy/utils/misc.py",
        "old_str": "...",
        "new_str": "...",
        "action_class":
          "StringReplaceArgs"
      },
      "observation": {
        "message":
          "The file has been edited...",
        "terminal": false
      }
    }],
    "visits": 15,
    "value": 917.0
  }]
}
\end{lstlisting}
\end{minipage}%
\hfill
\begin{minipage}[t]{0.32\textwidth}
\centering
\textbf{(c) Our Unified Schema}
\begin{lstlisting}[
  language=json,
  basicstyle=\ttfamily\tiny,
  frame=single,
  numbers=none,
  breaklines=true,
  breakatwhitespace=true,
  linewidth=\linewidth,
  showlines=true,
  aboveskip=4pt
]
{
  "trajectory_id": "astropy-7166",
  "repo_name": "astropy",
  "agent_name": "OpenHandsSonnet4",
  "final_outcome": true,
  "steps": [{
    "step_index": 1,
    "action_command": "str_replace",
    "files_touched":
      ["astropy/utils/misc.py"],
    "tests_run": [],
    "edit_size": 3,
    "error_message": "",
    "observation":
      "The file has been edited...",
      ...
  }]
}
\end{lstlisting}
\end{minipage}
\caption{Comparison of raw trajectory formats and the unified schema. (a) OpenHands records agent actions as a flat list. (b) Moatless organizes its trajectory into a layered tree structure. (c) Our parsing pipeline converts these different formats into one standard layout, allowing every agent's data to be handled using a single schema}
\label{fig:unified_schema}
\end{figure*}

Different agent frameworks record their trajectory data in different ways. As shown in Figure \ref{fig:unified_schema}, OpenHands~\cite{2025openhands} logs actions as a flat list, whereas Moatless~\cite{moatlessSweSearch} nests them inside a search tree. To ease our downstream analysis, we implement a parsing pipeline that converts diverse formats into a single, unified JSON schema.

The parsing is fully automated and deterministic, using rule-based normalization of shell commands, pattern matching to detect test runs, and exit-code to extract errors, without manual labeling or model inference. For error extraction, each step's observation is scanned for an exit code. An exit code of 0 means no error. For any other exit code, the parser records the Python traceback as the error message. All parsing rules and the analysis scripts are included in our replication package.\footnote{\url{www.github.com/SAILResearch/swe-agent-subset-selection}\label{fn:replication}} We confirm that parsed trajectories faithfully reflect the originals in two steps. First, we automatically compare every one of the parsed trajectories against its raw log: each recorded action and observation must appear verbatim in the log, each extracted exit code must match the one the log reports. All trajectories pass. We also manually verify 20 trajectories, reading the raw logs directly and confirming that every step is captured and that commands, test runs, and errors are classified correctly. All 20 matched. The unified schema captures what the agent does, not how the framework formats it, as illustrated in Figure \ref{fig:unified_schema}(c).

\subsection{Sanitize Trajectories}
\label{subsec:sanitization}
A central challenge in using trajectories for subset selection is data leakage. Outcome tokens, such as \texttt{passed} or \texttt{exit code 0} reveal whether a run succeeded, while repository names and test-harness artifacts can reveal task identity. If retained, such signals cause embeddings to be clustered by outcome or repository rather than by agent behavior. To avoid these unwanted signals, we develop a four-phase sanitization pipeline.

\subsubsection{Phase 1: Discover Corpus}
\label{subsubsec:sanitization1}
As we want our downstream analysis to reflect actual
agent behavior rather than formatting artifacts, we first analyze all tokens (whitespace-separated words, symbols, and numbers) across all trajectories to find tokens that are uninformative for comparing trajectories. These tokens fall into two categories: random identifiers, which almost never repeat across trajectories, and repetitive boilerplate, which appears identically in nearly all of them. We detect each with an automated statistical pass over the corpus, described below. This pass proposes candidate tokens and patterns, which we review once to build a stop-word list and a set of regex rules. We perform this calibration once per dataset, since the noise tokens are dataset-specific, while the procedure itself is general.

To detect \textbf{random identifiers} such as UUIDs, commit hashes, and temporary folder names, our analysis script converts each token into a character pattern: every letter in the token becomes \texttt{c} and every digit becomes \texttt{d} (e.g., \texttt{3f8a2b} $\rightarrow$ \texttt{dcdcdc}). Tokens that produce the same pattern form a group. For each pattern, the script counts how often it appears and how many distinct tokens produce it. A pattern produced by many distinct tokens is auto-generated text, such as a random seed or a temporary path, while a pattern produced by a few tokens repeated often is a genuine vocabulary term. The script outputs the list of patterns it flags as auto-generated. We then manually review the list of flagged patterns once per dataset and encode each confirmed pattern as a regular-expression rule that replaces matching tokens with a typed placeholder (for example, a temporary path becomes \texttt{[PATH]} and a random seed becomes \texttt{[RANDOM\_SEED]}), applied automatically across the corpus. Table~\ref{tab:sanitization_examples} shows examples.

\begin{table}[!t]
\centering
\caption{How we tell random identifiers from real tokens. For each token we
compare how often it appears (\#Occurrences) with how many \emph{different}
strings produce it (Distinct). A random identifier is a new string almost
every time, so the two counts match, and we replace the random identifier with a typed
placeholder. Boilerplate is one string repeated constantly, so it appears
very often but as a single distinct string, and we remove these boilerplate strings.}
\label{tab:sanitization_examples}
\small
\begin{tabular}{@{}llrrl@{}}
\toprule
& Example token & \#Occurrences & \#Distinct & Action \\
\midrule
\multirow{3}{*}{\rotatebox{90}{Identifier}}
& \texttt{PYTHONHASHSEED=1868256478}        & 491 & 491 & mask \texttt{[RANDOM\_SEED]} \\
& \texttt{/workspace/django\_\_django\dots} & 278 & 278 & mask \texttt{[PATH]} \\
& \texttt{sphinx/locale/da/\dots/sphinx.po} & 422 & 422 & mask \texttt{[PATH]} \\
\midrule
\multirow{3}{*}{\rotatebox{90}{Boiler.}}
& \texttt{execute\_bash:} & 79{,}431 & 1 & remove \\
& \texttt{[Command}       & 78{,}885 & 1 & remove \\
& \texttt{interpreter:}   & 73{,}076 & 1 & remove \\
\bottomrule
\end{tabular}
\end{table}

To detect \textbf{repetitive boilerplate}, we rank every token in the corpus by frequency. Zipf's
law~\cite{Zipf49, manning1999foundations} states that in natural language, token frequency is inversely proportional to rank: the most frequent token appears roughly twice as often as the second, three times as often as the third, and so on. The most frequent tokens in natural text are therefore common words like ``the'' or ``of.'' In our trajectories, however, the top-ranked tokens are not natural words but structural characters
(\texttt{{[}}, \texttt{\&\&}), harness commands (e.g., \texttt{pytest}, \texttt{make test}), and fixed environment paths (e.g., \texttt{/workspace}, \texttt{/tmp}) that repeat identically across nearly every trajectory. The script flags the top-ranked tokens, which we manually review once per dataset to compile a stop-word list. The list is then applied automatically to every trajectory during sanitization.

\subsubsection{Phase 2: Detect Outcome Leakage}
\label{subsubsec:sanitization2}
Phase~1 removes formatting noise, but a subtler problem
remains: some words in the trajectories directly reveal
whether a run has passed or failed. SWE-agent benchmarks score each run by whether the instance was resolved, so every trajectory already carries a pass/fail outcome. Words that reveal that outcome therefore add no new information, if left in the text,
may dominate the resulting embeddings, causing trajectories to be grouped by outcome rather than by the agent's actual problem-solving behavior.

To identify outcome-leakage words, we split the corpus into two groups (passing and failing trajectories) and count how often each word appears in each group. We then apply log-likelihood keyness analysis ($G^2$)~\cite{dunning1993accurate}, a statistical test that measures whether a word appears disproportionately more often in one group than in the other. A high $G^2$ score means the word is strongly associated with one group (i.e., test outcome). We use a significance threshold of $G^2 > 10.83$ ($p < 0.001$) to flag words for review. 

We then review the flagged words once per dataset. We remove a word if it reveals the instance outcome or the repository identity, while we keep a word if it reflects what the task requires the agent to do. Some calls are clear, such as removing ``passed'' or repository tooling like \texttt{pylint}. Others require judgment, so we treat this as a per-dataset calibration rather than a fixed rule. We group the flagged words as follows:

\begin{itemize}
\item \textbf{Removed.} Outcome indicators such as ``passed'' or ``exit code 0,'' and repository tooling such as \texttt{dvc} (Data Version Control) or \texttt{pylint} (a code linter), which are hard-coded into specific repositories' test harnesses. Keeping ``pylint'' would let the embedding associate it with failure, encoding which repository the agent is in rather than how it solves problems.
\item \textbf{Retained.} Terms that name what the task requires, such as \texttt{asyncio} (asynchronous concurrency) or \texttt{hypothesis} (property-based testing~\cite{maciver2019hypothesis}). These reflect the task's demands and the agent's approach, so they help distinguish trajectories by behavior and stay in the embeddings.
\end{itemize}

\subsubsection{Phase 3: Sanitize \& Mask}
\label{subsubsec:sanitization3}
We remove the outcome-leakage words identified in Phase 2, such as \texttt{passed} and \texttt{exit code 0}, and we automatically mask repository identities by replacing repository-specific terms with \texttt{[REPO\_VAR]}. We map structured error types to abstract tokens, for example \texttt{AssertionError} $\rightarrow$ \texttt{[ERR\_ASSERT]}, to preserve error-category information without retaining message-level details.

\subsubsection{Phase 4: Validate}
\label{subsubsec:sanitization4}
To verify whether our sanitization has removed outcome leakage
while preserving behavioral signal, we convert each trajectory's text into a numerical vector using TF-IDF~\cite{salton1988term}, a standard method that assigns higher weight to words that are distinctive to a document relative to the full corpus. We train a logistic regression classifier optimized with stochastic gradient descent~\cite{bottou2010sgd} on the datasets used in our empirical study (Section~\ref{subsec:data_points}) to predict whether each trajectory passed or failed. A logistic regression classifier learns which words are most predictive of an outcome, so by inspecting the words with the largest coefficients, we can see what information the text still contains.

We run the test on both the raw and sanitized corpora of all three datasets.
Similar accuracy is expected, since the classifier can still predict from task-related words once the outcome words are removed. We confirm that classification accuracy remains in a similar range (68.9\% raw versus 65.7\% sanitized on Single-setup, 72.8\% versus 76.2\% on Multi-model, and 78.3\% versus 80.3\% on Multi-agent), but the words the classifier relies on change. On the raw corpora, the words with the largest coefficients include the outcome indicator ``passed'' (on Single-setup) and repository identifiers and repository-specific tooling, such as \texttt{scrapy\_\_scrapy} and \texttt{pylint}, on all three datasets. On the sanitized corpora, the outcome indicators and repository terms disappear, and the words with the largest coefficients are engineering terms such as ``scope'' and ``hypothesis,'' which reflect the nature of the task rather than its outcome or origin. The shift in the top-coefficient words confirms that the classifier no longer relies on explicit outcome words. As a final check, we automatically scan 200 randomly sampled sanitized trajectories from each dataset for the words flagged for removal in Phases~1 and~2 and find no remaining instances.

\subsection{Vectorize Trajectories}
\label{subsec:approach_vectorization}
Our subset selection operates in a geometric space because trajectories that are close in this space reflect similar agent behavior. By measuring these distances, we aim to pick a subset that covers the full range of behaviors rather than just matching outcomes. To place trajectories in the geometric space, we convert each trajectory step's text into a numerical representation. We use Nomic Embed v1.5~\cite{nussbaum2024nomicembed}, a text embedding model that reads a piece of text, regardless of its length, and produces a fixed-size vector of embeddings that captures its meaning. Text that is similar in meaning produces vectors that are close together, allowing us to measure similarity between trajectory steps numerically. We select this model for two reasons: its 8,192-token context window can capture long error tracebacks without truncation, and it remains computationally practical on consumer hardware. We fix the embedding dimensionality at $d = 768$ using Matryoshka Representation Learning~\cite{kusupati2024matryoshka}. For each step, the embedding input is the step's sanitized action and observation from the unified schema (Figure~\ref{fig:unified_schema}(c)), joined into a single string. Other schema fields, such as the files touched or the edit size, are not part of the embedding input.

We compute two embedding variants:

\textbf{Pooled Embeddings:} Many subset selection methods require each instance to be represented as a fixed-dimensional vector. Each trajectory produces a matrix of $T \times 768$,  where $T$ is the number of steps the agent took (which is different across trajectories). To convert the matrix into a single vector, we apply statistics
pooling~\cite{snyder2018xvectors, Okabe2018AttentiveSPA}, a technique originally developed for speaker verification where audio recordings of different lengths must be compressed into a single fixed-length representation. According to pooling, we need to compute the mean and standard deviation of the $T$ step embeddings. The mean vector captures the central tendency of the trajectory's semantic content, while the standard deviation vector captures how much the agent's behavior varies from step to step. Okabe et al.~\cite{Okabe2018AttentiveSPA} show that including the standard deviation along with the mean captures long-term variation that mean alone would miss.

However, mean pooling discards chronological order. Mean pooling, although effective for capturing global semantics~\cite{reimers2019sentencebert}, is order-invariant and cannot distinguish between trajectories that differ only in the sequence of events. To encode temporal progression explicitly, we also calculate trajectory step embeddings for the beginning (step $0$), middle (step $\lfloor T/2 \rfloor$), and end (step $T-1$) of each trajectory. These structural anchors represent the agent's initial state, mid-execution behavior, and final state. We then concatenate all five vectors (mean, standard deviation, first step, middle step, and last step), producing a fixed-size vector of dimension $5 \times 768 = 3{,}840$ for each trajectory. This design combines statistical aggregation with temporal anchors.

\textbf{Time-Series Embeddings:} In contrast to pooled embeddings, this technique preserves the full sequence of step embeddings as a $T \times 768$ matrix, retaining the order in which the agent took its actions.

\subsection{Select Subset}
\label{subsec:approach_selection}
The developer specifies a subset size (e.g., 10\% of the full test suite). We consider two families of selection methods: baseline test selection, which draws instances at random, and geometric test selection, which selects instances deterministically from the embedding space. The baseline methods serve as comparison points in our evaluation, our proposed approach is the geometric selection of Section~\ref{subsubsec:geometric_selection}.

\subsubsection{Baseline Test Selection}
\label{subsubsec:baseline_selection}
We consider six baseline methods:
\begin{enumerate}
\item \textbf{Uniform Random (\textsc{Random}):} Picks instances at random from the full test suite, with no grouping or prior knowledge.
\item \textbf{Repository Stratified (\textsc{Repo-Strat}):} Groups instances by which repository they belong to (e.g., all Django issues together, all Flask issues together), mapping each instance to its source subsystem, then randomly samples from each group in proportion to its size. When a grouping produces a single group, for example if all instances come from one repository, the stratified method reduces to uniform random sampling within that group.
\item \textbf{Difficulty Stratified (\textsc{Diff-Strat}):} Splits instances into two groups based on historical outcomes: those the agent solved in the majority of the prior full runs (easy instances), and those it did not (hard instances). Then it randomly samples from each group in proportion to its size. Difficulty stratification requires at least one prior full run.
\item \textbf{Consistency Stratified (\textsc{Consist-Strat}):} Groups instances by their exact number of passes across prior runs. For example, with three prior runs, instances are separated into four groups: those that passed zero, one, two, or all three times. Grouping by consistency captures not just whether an instance is hard or easy, but how reliably the agent solves it across previous runs. With a single prior run, the pass counts reduce to the pass/fail split of difficulty stratification, so consistency stratification is applicable only when at least two prior runs are available.
\item \textbf{Repository $\times$ Difficulty (\textsc{Repo$\times$Diff}):} Combines repository grouping with difficulty grouping, creating subgroups (e.g., hard Django instances, easy Django instances) before sampling proportionally.
\item \textbf{Repository $\times$ Consistency (\textsc{Repo$\times$Consist}):} Combines repository grouping with consistency grouping, creating subgroups by repository and exact pass count before sampling proportionally.
\end{enumerate}
All stratified methods preserve each group's proportion of the full test suite when building the subset~\cite{cochran1977sampling, neyman1934representative}. For example, if a group contains 60\% of all instances, it receives 60\% of the subset's slots.

\subsubsection{Geometric Test Selection}
\label{subsubsec:geometric_selection}
The stratified baselines fix each group's share of the subset but still draw instances at random within each group. Our geometric methods keep the same grouping by pass counts (consistency groups, or the pass/fail split when only one prior full run is available) and replace the random draw within each group: instances are selected deterministically based on their position in the embedding space. We refer to
this combination of outcome grouping and geometric selection as
\textit{Embedding-Within-Strata}. We use five geometric algorithms:
\begin{enumerate}
\item \textbf{Centroid~\cite{manning2008introduction}:} Computes the average embedding of all instances in the group (the set of instances with the same pass count, as defined above), then selects the instances whose embeddings are closest to the average, up to the group's share of the subset. The selected instances are those where the agent followed the most common pattern of actions in the group, such as similar sequences of edits, test runs, and error handling.
\item \textbf{Facility Location~\cite{nemhauser1978analysis}:} Selects instances one at a time, each time choosing the instance that increases the total embedding similarity between the selected set and all remaining instances the most. This produces a set where every unselected instance has at least one similar representative in the subset.
\item \textbf{Medoid (PAM Build)~\cite{kaufman1990finding}:} First selects the single instance with the smallest total distance to all others (the most central real data point). It then adds the remaining instances one at a time, each time picking the one that best reduces the remaining distances to the already-selected set. Unlike Centroid, which computes an average point that may not correspond to any real instance, Medoid always picks actual instances from the data.
\item \textbf{Kennard-Stone~\cite{kennard1969computer}:} Selects instances one at a time, each time selecting the instance farthest from all previously selected instances. This maximizes diversity by spreading selections as far apart as possible in the embedding space.
\item \textbf{Core/Edge~\cite{manning2008introduction, kennard1969computer}:} A hybrid that allocates 90\% of the group's share using Centroid (picking typical instances) and 10\% using Kennard-Stone (picking the most different outliers). This balances representativeness with coverage of unusual behaviors.
\end{enumerate}

Given the same prior runs, embeddings, and subset size, each geometric algorithm always produces the same subset, with no random draws. On later runs, the developer evaluates only the selected subset and uses its resolve rate to estimate the full test suite's resolve rate.

%% file: Sections/Setup.tex
\section{Evaluation Setup}
\label{sec:eval_setup}

\subsection{Dataset: Three Agent Evaluation Scenarios}
\label{subsec:data_points}
To evaluate the effectiveness of our test selection approach (Section~\ref{sec:methodology}) under different types of agent changes, we curate a dataset comprising three evaluation scenarios: (1) reruns of the same agent configurations, (2) changes to the model or configurations of a given agent, and (3) changes to the agent framework. We collect trajectory data from established benchmarks to represent each of these scenarios:

\begin{enumerate}
    \item \textbf{Single-setup: Rerun Same Agent.} This is a sanity-check scenario where nothing changes between runs. The same agent is run repeatedly in the same deployment environment, so the only variation comes from the stochasticity of the model's outputs. This scenario captures the easiest case, where a subset only needs to absorb run-to-run noise rather than any real change in behavior. We use existing, publicly available trajectories of the OpenHands agent with the Qwen3-Coder model, which we downloaded from SWE-Rebench~\cite{trofimova2025openhandstrajs}\footnote{\url{https://huggingface.co/datasets/nebius/SWE-rebench-openhands-trajectories}}. The full dataset contains 67,074 trajectories over 6,306 instances, with each instance rerun a varying number of times. We keep instances rerun 5 to 10 times, a range comparable to the run counts of the other datasets included in the study. The retained dataset includes 45 runs over the SWE-Rebench benchmark instances (3,188 instances, 25,279 trajectories). The runs form six independent groups, where, in each group, the same set of instances is rerun a fixed number of times, from 5 reruns for one group up to 10 for another. The groups differ in size, so the total number of trajectories is not a simple product of instances and runs.
    \item \textbf{Multi-model: Change Models.} This scenario captures a common development case where the agent framework stays the same but the underlying foundation model or its execution settings change. A model change can alter how instances are solved and which of them pass, so the subset must remain representative as per-instance outcomes change between runs. Unlike the Single-setup dataset, we downloaded OpenHands trajectories from the SWE-bench experiments repository\footnote{\url{https://github.com/SWE-bench/experiments}}, where each run pairs OpenHands with a different model and diverse execution settings (e.g., inference scaling, adjusted reasoning effort, and varying iteration limits). This dataset consists of 6 runs of 500 SWE-bench Verified instances each, yielding 3,000 trajectories.
    \item \textbf{Multi-agent: Change Agent Framework.} To assess subset selection under more substantial changes to the agent's architecture, we include trajectories from five agent frameworks: OpenHands, Moatless, Lingxi, Trae, and Refact, paired with models including Claude 3.5 Sonnet, Claude 4 Sonnet, Kimi K2, and multi-model ensembles. A framework change alters the agent's entire problem-solving strategy, not just how a fixed agent behaves, so the historical trajectories used for selection may no longer reflect the new agent's behavior. This is the hardest scenario for our approach and tests where trajectory-based selection reaches its limits. Also downloaded from the SWE-bench experiments repository, the dataset comprises 7 runs on 500 SWE-bench Verified instances, yielding 3,500 trajectories. 
\end{enumerate}

\subsection{Evaluation Protocol}
\begin{figure*}[!t]
\centering
\includegraphics[width=\linewidth]{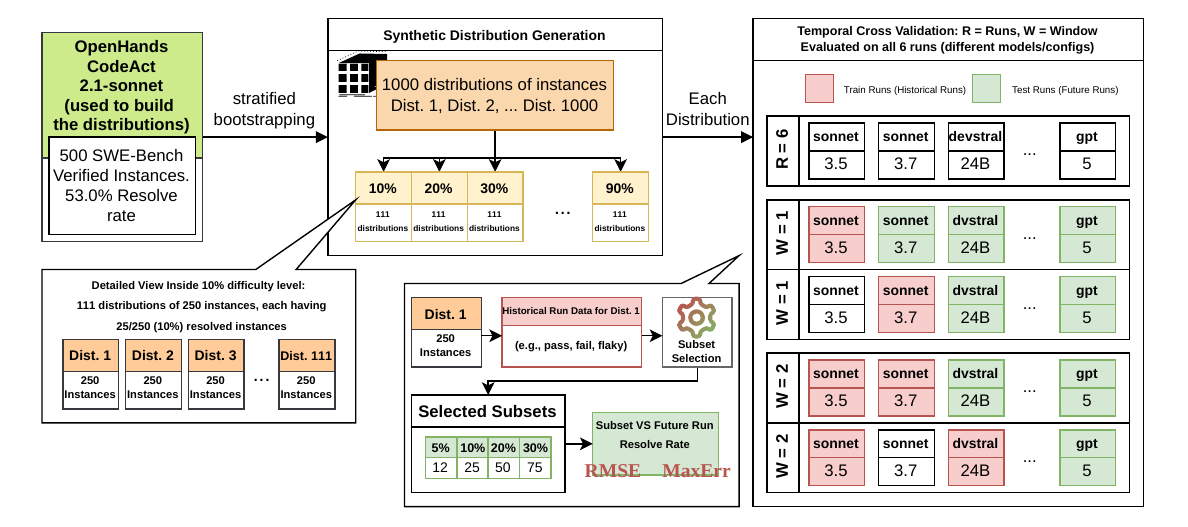}
\caption{\textbf{Evaluation protocol}, illustrated on the Multi-model dataset. The Multi-agent dataset follows the same protocol. The Single-setup dataset follows the same protocol without the synthetic distribution generation step, since its 45 empirical runs already provide enough data. Left: stratified bootstrapping over the source run generates 1,000 synthetic distributions, each a set of 250 benchmark instances, spread across nine difficulty levels. Right: each distribution is evaluated on all runs under temporal cross-validation, which splits runs into past (training) and future (test) runs. Example splits are shown for window sizes one and two. Bottom: the evaluation flow for one distribution, comparing the subset resolve rate against the full distribution's resolve rate on future runs to compute RMSE and MaxErr.}
\label{fig:evaluation_protocol}
\end{figure*}

Figure~\ref{fig:evaluation_protocol} provides a high-level overview of our evaluation protocol. Because the source datasets differ substantially in the number of runs they contain and the difficulty of the instances, the protocol has two main phases. First, we generate representative, synthetic trajectory distributions by sampling benchmark instances (across all runs), at controlled resolve rates (10\% through 90\%) to validate our methods across a broad range of difficulty levels. Second, we simulate realistic development conditions by treating earlier runs as historical data, from which we compute embeddings and select the subset, and later runs as unseen future runs, preserving chronological order and preventing look-ahead bias. The following sections describe these steps in detail.

\subsubsection{Synthetic Distribution Generation}
\label{subsec:synthetic_distribution}
The \textbf{Single-setup} dataset is a sanity-check scenario containing 45 empirical runs over 3,188 instances, which provide enough data for direct evaluation. In contrast, the \textbf{Multi-model} and \textbf{Multi-agent} datasets contain only 6 and 7 runs, respectively, which is insufficient for robust comparison across selection methods. We therefore generate synthetic distributions only for the Multi-model and Multi-agent datasets, applying the same procedure to each. Everything after this distributions generation step applies identically to all three datasets.

To support evaluation over a broad range of test suite conditions for the Multi-model and Multi-agent datasets, we sample 1,000 synthetic test suite distributions from the 500-instance SWE-Bench Verified pool, where each distribution is a set of benchmark instances. We build the distributions from a single run of each dataset, which we call the source run. For both the Multi-model and Multi-agent datasets, the source run is the earliest run (OpenHands CodeAct 2.1-sonnet, 53\% resolve rate), whose near-balanced resolve rate allows sampling distributions at both low and high difficulty. The cross-run validity analysis at the end of this section verifies that the difficulty levels defined on the source run remain valid on all other runs. Each distribution contains $N = 250$ instances sampled without replacement~\cite{cochran1977sampling}. For each of its instances, a distribution carries the instance's trajectories across all runs of the dataset (e.g., $500 \times 6 = 3{,}000$ trajectories in the Multi-model dataset). To ensure balanced coverage, we control the sampling so that distributions are spread evenly across nine difficulty levels, defined by their resolve rate on the source run (10\%, 20\%, \dots, 90\%), producing approximately 111 distributions per level~\cite{efron1993introduction}. We refer to these levels as source-run difficulty levels throughout the paper. Having many distributions at each level allows us to compare methods using rank-based statistical tests that make no assumptions about how errors are distributed~\cite{arcuri2011practical}, rather than relying on tests that assume a specific error distribution, such as a normal distribution.

Selection methods that involve randomness, such as uniform random sampling and the stratified sampling baselines, can produce a different subset on every draw. For each distribution and train/test split, we therefore keep the distribution fixed and repeat the subset selection with 500 independent random seeds, reporting results across all 500 subsets. To check that 500 seeds are enough, we repeated the experiment with 1,000 and 2,000 seeds. The method rankings and effect sizes do not change, so we fix the number of seeds at 500 for computational efficiency.

To verify that the synthetic distributions are sufficiently diverse, we compute the Jaccard similarity~\cite{jaccard1901distribution} of every pair of the 1{,}000 distributions: the number of instances two distributions share, divided by the number of distinct instances they cover together.
The mean over all pairs is 0.35, which is as diverse as the design allows. With 1,000 distributions of 250 instances drawn from a pool of 500, the average pairwise overlap is about 125 instances by construction, corresponding to a mean Jaccard similarity of approximately 0.33. The observed mean similarity of 0.35 is therefore close to the minimum average overlap allowed by the design. The small gap above the floor comes from the stratified sampling: distributions at the extreme difficulty levels must draw most of their instances from the same outcome group (a 10\% distribution takes 225 of the 235 instances that fail on the source run), so they overlap more than an unconstrained draw would. Within the middle difficulty levels (40\% to 60\% source resolve rate), where this constraint is weakest, the mean similarity is 0.33 to 0.36, at the floor. The overlap is therefore a property of the pool size rather than of the sampling, and Sections~\ref{subsec:statistical_analysis} and~\ref{subsec:conclusion_validity} state how it affects the statistical tests.

\paragraph{Cross-run validity of difficulty levels.}
\label{subsec:cross_run_validity}
We estimate all difficulty levels from the single source run to avoid the cost of evaluating every candidate distribution on every run. Using one run is valid because the difficulty ranking holds across runs: distributions ranked harder on the source run stay harder on the remaining runs, even though the absolute resolve rates shift as models get stronger (e.g., a distribution at the lowest difficulty level resolves at 10\% on the source run but at 42\% under Sonnet 3.7). Section~\ref{sec:threatstovalidity} provides the supporting correlation analysis.

\begin{table}[t]
\caption{Temporal cross-validation splits, illustrated for the Multi-model dataset ($R{=}6$ runs) at window size $W{=}2$. The other datasets and window sizes follow the same enumeration.
\textbf{Train} (2 runs): the subset selection method uses outcomes and trajectories from these runs
to compute outcome groups, calculate the trajectories embeddings, and select the subset.
\textbf{Test} (runs strictly after the latest training run): the selected subset is evaluated on these runs
by comparing its resolve rate against the full distribution's resolve rate on the run under test.
Runs that fall before or between the training runs are unused (---).
Of the $\binom{6}{2} = 15$ possible combinations, 5 have the latest training run at GPT-5 (run~6),
leaving no runs for evaluation, these are excluded, yielding 10 usable splits.
This procedure is applied independently to each of the 1,000 synthetic distributions,
for stochastic methods, each split is repeated with 500 random seeds.}
\label{tab:tcv-splits}
\centering
\small
\begin{tabular}{c cccccc c}
\toprule
 & \multicolumn{6}{c}{\textbf{Runs (chronological order)}} & \\
\cmidrule(lr){2-7}
\textbf{Split}
 & \rotatebox{55}{Sonnet 3.5}
 & \rotatebox{55}{Sonnet 3.7}
 & \rotatebox{55}{Devstral}
 & \rotatebox{55}{Sonnet 4}
 & \rotatebox{55}{Kimi K2}
 & \rotatebox{55}{GPT-5}
 & \textbf{\# Test} \\
\midrule
1  & \cellcolor{trainbg}Train & \cellcolor{trainbg}Train & \cellcolor{testbg}Test & \cellcolor{testbg}Test & \cellcolor{testbg}Test & \cellcolor{testbg}Test & 4 \\
2  & \cellcolor{trainbg}Train & ---                      & \cellcolor{trainbg}Train & \cellcolor{testbg}Test & \cellcolor{testbg}Test & \cellcolor{testbg}Test & 3 \\
3  & \cellcolor{trainbg}Train & ---                      & ---                      & \cellcolor{trainbg}Train & \cellcolor{testbg}Test & \cellcolor{testbg}Test & 2 \\
4  & \cellcolor{trainbg}Train & ---                      & ---                      & ---                      & \cellcolor{trainbg}Train & \cellcolor{testbg}Test & 1 \\
5  & ---                      & \cellcolor{trainbg}Train & \cellcolor{trainbg}Train & \cellcolor{testbg}Test & \cellcolor{testbg}Test & \cellcolor{testbg}Test & 3 \\
6  & ---                      & \cellcolor{trainbg}Train & ---                      & \cellcolor{trainbg}Train & \cellcolor{testbg}Test & \cellcolor{testbg}Test & 2 \\
7  & ---                      & \cellcolor{trainbg}Train & ---                      & ---                      & \cellcolor{trainbg}Train & \cellcolor{testbg}Test & 1 \\
8  & ---                      & ---                      & \cellcolor{trainbg}Train & \cellcolor{trainbg}Train & \cellcolor{testbg}Test & \cellcolor{testbg}Test & 2 \\
9  & ---                      & ---                      & \cellcolor{trainbg}Train & ---                      & \cellcolor{trainbg}Train & \cellcolor{testbg}Test & 1 \\
10 & ---                      & ---                      & ---                      & \cellcolor{trainbg}Train & \cellcolor{trainbg}Train & \cellcolor{testbg}Test & 1 \\
\bottomrule
\end{tabular}
\end{table}

\subsubsection{Temporal Cross-Validation}
In practice, a developer selects a subset based on runs that have already been completed and uses it to evaluate future runs that have not yet
occurred~\cite{kim2002history, elbaum2014}. To replicate this setting, we ensure that subset selection (Section~\ref{subsec:approach_selection}) is trained only on past data, using it to compute embeddings, the pass counts that form the outcome groups, and the selected subset, and is always tested on later runs it has never seen. Given $R$ runs (each covering the same set of instances) sorted chronologically and a window size $W$, we enumerate all $\binom{R}{W}$ combinations rather than only the most recent $W$ runs. The enumeration is a stress test of the
selection methods across different histories rather than a claim that a developer would assemble a history this way. Splits whose training runs are not contiguous place the
training runs further from the test run than a deployment history would, so they probe the methods under older history rather than under the most favorable conditions. We refer to each train/test partition as a \textit{temporal split}. As an illustration, Table~\ref{tab:tcv-splits} lists all 10 usable splits for $W = 2$ on the Multi-model dataset ($R = 6$). The same enumeration applies to every dataset and window size, with the number of splits determined by $R$ and $W$. We evaluate every window size from $W = 1$ to $W = R - 1$ for each dataset. Runs that fall between the two training runs are skipped: for example, split~2 trains on Sonnet~3.5 and Devstral, skipping Sonnet~3.7 because it falls between them, and tests on Sonnet~4 through GPT-5.

\subsection{Evaluation Metrics}
\label{subsec:metrics}
We evaluate subset quality using two main metrics: Root Mean Square Error and Max Error.

\paragraph{Root Mean Square Error (RMSE)}
Our primary metric is RMSE, which measures how closely the subset tracks the full population's resolve rate across unseen future runs~\cite{shepperd2012evaluating}:
\[
\text{RMSE} = \sqrt{\frac{1}{|\mathcal{T}|} \sum_{r \in \mathcal{T}} 
\left(\overline{y}_{S,r} - \overline{y}_{P,r}\right)^2}
\]
Here, $S$ denotes the selected subset; $P$ denotes the full population; and $\overline{y}_{\cdot,r}$ denotes the resolve rate on run $r$, computed over either the subset ($S$) or the full population ($P$). The set $\mathcal{T}$ contains the held-out test runs that occur chronologically after the training window. The subset $S$ is selected using the training runs only. On each test run, we then measure the resolve rate of the subset's instances and compare it against the full population's resolve rate on that same run, which serves as the ground truth. For deterministic methods, which produce one fixed subset per temporal split, we report the mean RMSE across all temporal splits for each distribution. For stochastic methods, each temporal split is repeated with 500 independent random seeds, each producing a different subset, and we report the mean RMSE across all seeds and splits.

\paragraph{Maximum Error (MaxErr)}
RMSE averages errors across runs and can obscure worst-case risk~\cite{shepperd2012evaluating}. Hence, we also report Maximum Error (MaxErr), computed per distribution and expressed as a percentage. For a given distribution, MaxErr is the largest absolute difference in the resolve rate between the selected subset and the full population, taken over all held-out test runs and all temporal splits:

\[
\text{MaxErr}_k = 100 \times \max_{s \in \mathcal{S}} \max_{r \in \mathcal{T}_s} \left| \overline{y}_{S_k,r}^{(s)} - \overline{y}_{P,r}^{(s)} \right|
\]
Here, $\mathcal{S}$ denotes the set of temporal splits for a given distribution, $\mathcal{T}_s$ denotes the held-out test runs for split $s$, and $S_k$ denotes the subset selected with random seed $k$. For stochastic methods that draw 500 random subsets, each seed $k$ therefore produces its own $\text{MaxErr}_k$. We report three summary statistics: the median (typical draw), the 95th percentile (P95, a bad-luck draw), and the maximum over all seeds (worst). For deterministic methods that produce a single fixed subset, all three coincide. Minimizing MaxErr is important in regression testing settings, where a single misleading estimate can lead to an incorrect deployment decision.

\subsection{Statistical Analysis}
\label{subsec:statistical_analysis}
\paragraph{Significance Testing}
To determine whether one subset selection approach performs significantly better than another, we use the Wilcoxon signed-rank test~\cite{wilcoxon1945individual}, which compares two methods across the same set of instance distributions and checks whether one consistently produces lower error than the other. Unlike tests that assume errors to follow a normal statistical distribution, this test only relies on the relative ranking of differences, making it suitable for our setting. Because we run many pairwise comparisons, we apply Holm-Bonferroni correction~\cite{holm1979simple} to the p-values within each comparison family, defined as all method pairs compared on the same dataset and subset size. Holm's step-down procedure sorts the $m$ p-values of a family and compares the smallest against $\alpha/m$, the next against $\alpha/(m-1)$, and so on, at $\alpha = 0.05$.

We also report Dem\v{s}ar's paired dominance ranking~\cite{demsar2006statistical}. The dominance ranking assigns each method a single score: the number of other methods that significantly outperform it after Holm correction. A method with rank 0 is not beaten by any other method. The synthetic distributions are drawn from a pool of 500 instances and overlap (mean Jaccard 0.35, Section~\ref{subsec:synthetic_distribution}), so the paired samples are not independent as the test assumes. Dependence inflates the effective sample size and makes $p$-values optimistic, and reporting effect sizes does not correct the $p$-values. We therefore rest our conclusions on Cliff's $\delta$, treat a negligible $\delta$ as no difference regardless of the $p$-value, and report a robustness check on the dependence in Section~\ref{subsec:conclusion_validity}.

\paragraph{Effect Size}
While the Wilcoxon test tells us whether a difference exists, it does not tell us how large that difference is. To measure the difference, we report Cliff's $\delta$~\cite{cliff1993dominance}, which quantifies how often one method outperforms another across all paired comparisons. Because it is based purely on counting wins and losses, it makes no assumptions about how the data are distributed. We interpret effect sizes using the standard thresholds: negligible ($|\delta| < 0.147$), small ($|\delta| < 0.33$), medium ($|\delta| < 0.474$), and large ($|\delta| \geq 0.474$)~\cite{romano2006appropriate}.

%% file: Sections/Results.tex
\section{Experiment Results}
\label{sec:results}

\input{Sections/RQ1}

\input{Sections/RQ2}

\input{Sections/RQ3}

\input{Sections/RQ4}

%% file: Sections/RQ1.tex
\subsection{RQ1: How effective are baseline subset selection approaches at approximating full-test suite resolve rates?}
\label{sec:results_rq1}

\subsubsection*{\textbf{Motivation:}} Before testing our embedding-based methods, we establish baselines by evaluating how well simple approaches work on their own. If random sampling or grouping by past outcomes already produces reliable subsets, a more complex method would not be needed.

\subsubsection*{\textbf{Approach:}} 
We evaluate the six baseline selection methods described in Section~\ref{subsubsec:baseline_selection}. Each method is run with 500 independent random seeds under our full evaluation protocol, reporting Root Mean Square Error (RMSE) and Maximum Error (MaxErr) against the resolve rate of all instances in the distribution. 

In this RQ, we evaluate how performance changes when we adjust the number of historical runs available ($W$), the difficulty level of the test suite (nine difficulty levels, Section~\ref{subsec:synthetic_distribution}), and the size of the selected test subset (from 5\% to 30\% of the full test suite).

\subsubsection*{\textbf{Finding 1.1:}} \textbf{Consistency Stratification is the strongest baseline}
\label{sec:rq1_finding1}
Among the six baseline approaches, a performance hierarchy appears across experimental settings. Methods that use historical outcome information outperform uniform random sampling, with \textsc{Consist-Strat} (Consistency Stratification) emerging as the strongest baseline overall. It achieves the lowest median RMSE in all datasets and subset sizes (Table~\ref{tab:rq1-naive}). It reduces median RMSE relative to Random by 51\% in \textbf{Single-setup}, by 31--32\% in \textbf{Multi-model}, and by 40--42\% in \textbf{Multi-agent}. For example, at 5\% subset size, it reduces RMSE from 0.07 to 0.04 (\textbf{Single-setup}), from 0.11 to 0.07 (\textbf{Multi-model}), and from 0.10 to 0.06 (\textbf{Multi-agent}). The same holds at the 95th percentile of RMSE across distributions (P95 in Table~\ref{tab:rq1-naive}), so its advantage is not limited to typical distributions but also holds where errors are the largest.

The results of \textsc{Consist-Strat} suggest that 
grouping instances by their exact pass count across prior runs captures how reliably an instance is resolved. \textsc{Random} can accidentally over-represent easy instances or under-represent brittle ones, skewing the estimated resolve rate. \textsc{Consist-Strat} preserves the distribution of always-passing, sometimes-passing, and always-failing instances, so the subset remains representative even when the underlying model changes. Under Dem\v{s}ar's paired dominance ranking, \textsc{Consist-Strat} significantly outperforms Random across all datasets and subset sizes ($p < 0.001$, $\delta = 1.000$, large effect size), with a perfect 1000/0/0 win/tie/loss record across the 1,000 synthetic evaluation distributions.

\begin{figure*}[t]
\centering
\includegraphics[width=\textwidth]{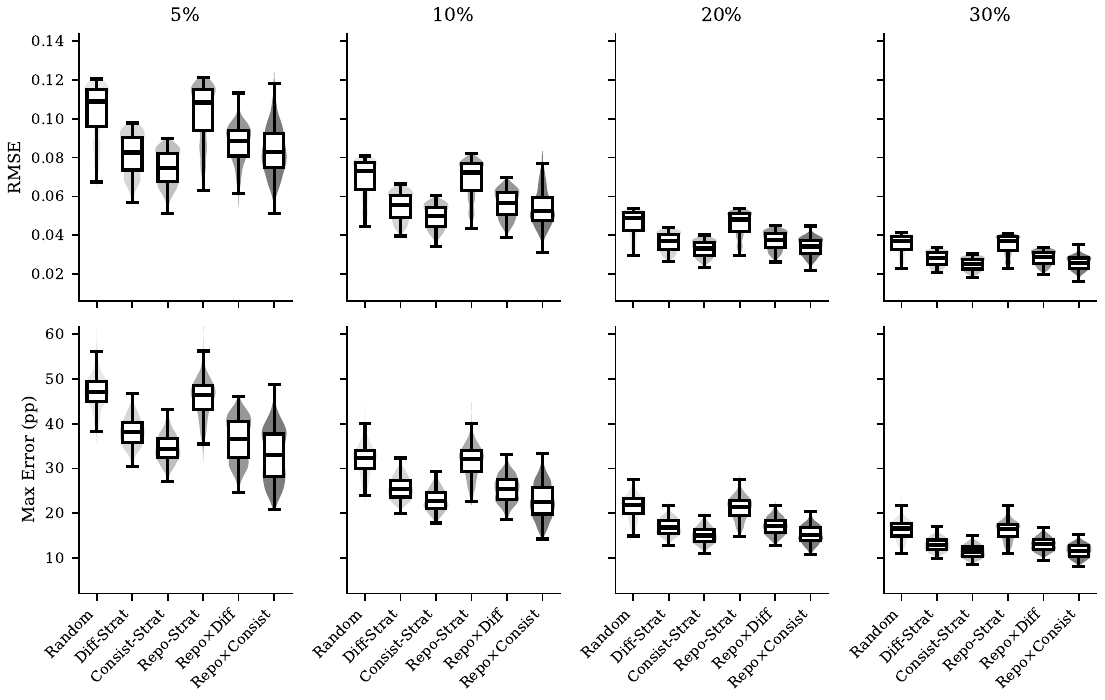}
\caption{Distributions of RMSE and Max Error for the baselines across 1,000 synthetic evaluations (Multi-model).
\textbf{Top Row:} Average predictive accuracy (RMSE),
\textbf{Bottom Row:} Per-seed worst-case error (MaxErr). Per-seed statistics are reported in Table~\ref{tab:rq1-maxerr}, revealing the severe risk inherent in all stochastic subset selection approaches. At a 5\% subset size, the best baseline's worst draw produces a worst-case error of 35\%, and even at 30\% its worst draw still produces 11\%.}
\label{fig:rq1-naive}
\end{figure*}

\begin{table*}[t]
\centering
\caption{RQ1: Baseline RMSE across datasets and subset sizes. Med = median RMSE, P95 = 95th percentile RMSE across $N{=}1{,}000$ synthetic distributions. The strongest baseline is \textbf{bolded}. MaxErr is reported separately in Table~\ref{tab:rq1-maxerr}.}
\label{tab:rq1-naive}
\resizebox{0.65\textwidth}{!}{
\begin{tabular}{llrrrrrr}
\toprule
Approach & Subset & \multicolumn{2}{c}{\textbf{Single-Setup}} & \multicolumn{2}{c}{\textbf{Multi-Model}} & \multicolumn{2}{c}{\textbf{Multi-Agent}} \\
\cmidrule(lr){3-4} \cmidrule(lr){5-6} \cmidrule(lr){7-8}
 &  & Med & P95 & Med & P95 & Med & P95 \\
\midrule
Random & 5\% & 0.0744 & 0.0777 & 0.1090 & 0.1182 & 0.1048 & 0.1183 \\
Diff-Strat &  & 0.0410 & 0.0474 & 0.0827 & 0.0954 & 0.0689 & 0.0839 \\
\textbf{Consist-Strat} &  & \textbf{0.0367} & 0.0416 & \textbf{0.0746} & 0.0868 & \textbf{0.0633} & 0.0764 \\
Repo-Strat &  & 0.0761 & 0.0769 & 0.1085 & 0.1183 & 0.1037 & 0.1171 \\
Repo$\times$Diff &  & 0.0983 & 0.1038 & 0.0886 & 0.0993 & 0.0803 & 0.1023 \\
Repo$\times$Consist &  & 0.0709 & 0.0905 & 0.0830 & 0.1050 & 0.0783 & 0.1171 \\
\midrule
Random & 10\% & 0.0511 & 0.0534 & 0.0730 & 0.0792 & 0.0720 & 0.0807 \\
Diff-Strat &  & 0.0283 & 0.0332 & 0.0556 & 0.0640 & 0.0470 & 0.0562 \\
\textbf{Consist-Strat} &  & \textbf{0.0250} & 0.0287 & \textbf{0.0498} & 0.0576 & \textbf{0.0422} & 0.0507 \\
Repo-Strat &  & 0.0470 & 0.0480 & 0.0723 & 0.0788 & 0.0706 & 0.0794 \\
Repo$\times$Diff &  & 0.0378 & 0.0497 & 0.0564 & 0.0661 & 0.0485 & 0.0570 \\
Repo$\times$Consist &  & 0.0340 & 0.0409 & 0.0526 & 0.0715 & 0.0443 & 0.0556 \\
\midrule
Random & 20\% & 0.0342 & 0.0356 & 0.0487 & 0.0527 & 0.0474 & 0.0532 \\
Diff-Strat &  & 0.0189 & 0.0221 & 0.0369 & 0.0426 & 0.0309 & 0.0372 \\
\textbf{Consist-Strat} &  & \textbf{0.0167} & 0.0191 & \textbf{0.0331} & 0.0385 & \textbf{0.0275} & 0.0335 \\
Repo-Strat &  & 0.0315 & 0.0324 & 0.0481 & 0.0525 & 0.0465 & 0.0527 \\
Repo$\times$Diff &  & 0.0169 & 0.0198 & 0.0377 & 0.0431 & 0.0314 & 0.0377 \\
Repo$\times$Consist &  & 0.0538 & 0.0632 & 0.0342 & 0.0400 & 0.0291 & 0.0349 \\
\midrule
Random & 30\% & 0.0260 & 0.0273 & 0.0371 & 0.0403 & 0.0361 & 0.0405 \\
Diff-Strat &  & 0.0145 & 0.0167 & 0.0281 & 0.0327 & 0.0236 & 0.0283 \\
\textbf{Consist-Strat} &  & \textbf{0.0128} & 0.0147 & \textbf{0.0253} & 0.0293 & \textbf{0.0208} & 0.0253 \\
Repo-Strat &  & 0.0305 & 0.0317 & 0.0368 & 0.0401 & 0.0355 & 0.0401 \\
Repo$\times$Diff &  & 0.0219 & 0.0234 & 0.0286 & 0.0329 & 0.0238 & 0.0286 \\
Repo$\times$Consist &  & 0.0385 & 0.0450 & 0.0259 & 0.0300 & 0.0217 & 0.0262 \\
\bottomrule
\end{tabular}}
\end{table*}

\begin{table*}[t]
\centering
\caption{RQ1: Per-seed worst-case error (MaxErr, \%) across datasets and subset sizes. Each of 500 random seeds selects one subset whose MaxErr is computed independently. Med = median across seeds (the error a typical draw produces), P95 = 95th percentile (an unlucky draw), Worst = maximum across all seeds. The strongest baseline is \textbf{bolded}.}
\label{tab:rq1-maxerr}
\resizebox{0.8\textwidth}{!}{
\begin{tabular}{llrrrrrrrrr}
\toprule
Approach & Subset & \multicolumn{3}{c}{\textbf{Single-Setup}} & \multicolumn{3}{c}{\textbf{Multi-Model}} & \multicolumn{3}{c}{\textbf{Multi-Agent}} \\
\cmidrule(lr){3-5} \cmidrule(lr){6-8} \cmidrule(lr){9-11}
 &  & Med & P95 & Worst & Med & P95 & Worst & Med & P95 & Worst \\
\midrule
Random & 5\% & 9.20 & 13.78 & 17.91 & 20.21 & 32.97 & 47.22 & 21.76 & 33.82 & 47.78 \\
Diff-Strat &  & 5.35 & 8.37 & 11.35 & 16.25 & 26.57 & 38.16 & 15.91 & 25.34 & 36.95 \\
\textbf{Consist-Strat} &  & \textbf{4.58} & 7.16 & 9.77 & \textbf{14.42} & 23.83 & 34.61 & \textbf{14.28} & 22.15 & 31.25 \\
Repo-Strat &  & 9.93 & 14.05 & 17.28 & 19.94 & 32.14 & 45.72 & 21.24 & 32.59 & 45.98 \\
Repo$\times$Diff &  & 10.46 & 12.49 & 14.14 & 16.89 & 26.31 & 36.52 & 16.97 & 24.51 & 34.15 \\
Repo$\times$Consist &  & 11.01 & 12.38 & 13.35 & 15.72 & 24.03 & 33.00 & 16.43 & 22.48 & 28.86 \\
\midrule
Random & 10\% & 6.32 & 9.60 & 12.99 & 13.76 & 22.36 & 32.12 & 15.03 & 23.21 & 32.85 \\
Diff-Strat &  & 3.92 & 5.79 & 7.76 & 10.97 & 17.92 & 25.60 & 10.86 & 17.04 & 24.68 \\
\textbf{Consist-Strat} &  & \textbf{3.35} & 4.99 & 6.51 & \textbf{9.63} & 15.87 & 22.92 & \textbf{9.60} & 14.94 & 21.23 \\
Repo-Strat &  & 7.24 & 10.12 & 12.78 & 13.62 & 22.07 & 31.50 & 14.75 & 22.68 & 31.92 \\
Repo$\times$Diff &  & 5.16 & 6.48 & 7.47 & 11.12 & 17.97 & 25.35 & 10.95 & 16.60 & 23.47 \\
Repo$\times$Consist &  & 5.16 & 6.06 & 6.63 & 10.28 & 16.41 & 22.68 & 9.90 & 14.66 & 19.66 \\
\midrule
Random & 20\% & 4.19 & 6.42 & 8.50 & 9.17 & 14.95 & 21.33 & 9.99 & 15.40 & 21.51 \\
Diff-Strat &  & 2.58 & 3.85 & 5.16 & 7.32 & 11.86 & 16.95 & 7.16 & 11.08 & 15.83 \\
\textbf{Consist-Strat} &  & \textbf{2.15} & 3.28 & 4.41 & \textbf{6.41} & 10.53 & 15.11 & \textbf{6.30} & 9.78 & 13.84 \\
Repo-Strat &  & 5.02 & 6.75 & 8.14 & 9.09 & 14.84 & 21.14 & 9.83 & 15.16 & 21.21 \\
Repo$\times$Diff &  & 4.55 & 5.28 & 5.86 & 7.42 & 12.03 & 17.11 & 7.20 & 11.08 & 15.65 \\
Repo$\times$Consist &  & 5.50 & 5.95 & 6.31 & 6.61 & 10.76 & 15.20 & 6.47 & 9.80 & 13.51 \\
\midrule
Random & 30\% & 3.22 & 4.90 & 6.50 & 7.01 & 11.43 & 16.30 & 7.61 & 11.73 & 16.28 \\
Diff-Strat &  & 1.99 & 2.92 & 3.77 & 5.57 & 9.06 & 12.90 & 5.44 & 8.43 & 11.94 \\
\textbf{Consist-Strat} &  & \textbf{1.64} & 2.45 & 3.28 & \textbf{4.89} & 8.02 & 11.44 & \textbf{4.78} & 7.43 & 10.50 \\
Repo-Strat &  & 4.42 & 5.65 & 6.68 & 6.95 & 11.32 & 16.04 & 7.49 & 11.55 & 16.03 \\
Repo$\times$Diff &  & 3.75 & 4.29 & 4.71 & 5.63 & 9.14 & 13.03 & 5.46 & 8.43 & 11.86 \\
Repo$\times$Consist &  & 3.56 & 3.88 & 4.08 & 5.02 & 8.17 & 11.53 & 4.90 & 7.45 & 10.34 \\
\bottomrule
\end{tabular}}
\end{table*}

\subsubsection*{\textbf{Finding 1.2:}} \textbf{Repository metadata is not a meaningful predictor of instance difficulty.}
\label{sec:rq1_finding2}
Repository Stratification (\textsc{Repo-Strat}) provides little value across all conditions and fails to separate instances by evaluation difficulty. Table~\ref{tab:rq1-naive} shows that in \textbf{Multi-model} at a 5\% subset size, \textsc{Repo-Strat} yields a median RMSE of 0.10, which is nearly identical to Random. The pattern persists at larger subset sizes. For example, in \textbf{Multi-agent} at 30\%, \textsc{Repo-Strat} improves on Random by less than 2\%. This seems to suggest that repository identity is a weak proxy for instance difficulty. In agent regression testing, evaluation difficulty is shaped more by how the agent interacts with the instance than by the repository which the instance belongs to.

The baseline approaches that combine repository metadata with outcome-aware strategies, for example \textsc{Repo$\times$Consist}, consistently worsen \textbf{average predictive accuracy (RMSE)} relative to the corresponding non-repository variants. In \textbf{Multi-model} at a 5\% subset size, \textsc{Repo$\times$Consist} yields a median RMSE of 0.08, compared with 0.07 for \textsc{Consist-Strat}, a degradation of 11\%. This pattern holds in nearly all subset sizes and datasets (Table~\ref{tab:rq1-naive}), suggesting that the added repository dimension over-fragments the test suite into strata that are too small for reliable proportional sampling (Table~\ref{tab:rq1-naive}: in \textbf{Single-Setup}, \textsc{Repo$\times$Consist} RMSE increases from 0.0340 at 10\% to 0.0538 at 20\%, then drops to 0.0385 at 30\%).

\subsubsection*{\textbf{Finding 1.3:}} \textbf{All stochastic approaches retain substantial worst-case risk, regardless of stratification quality.}
\label{sec:rq1_finding3}
Although \textsc{Consist-Strat} achieves the best average accuracy among baseline approaches, Figure~\ref{fig:rq1-naive} shows that all stochastic methods still carry substantial worst-case risk. The bottom row of the figure highlights MaxErr directly, revealing a long tail of severe mispredictions.

Table~\ref{tab:rq1-maxerr} reports per-seed MaxErr statistics. At 5\%, a typical random draw (Med) already produces a worst-case error of 20.21\% in \textbf{Multi-model} and 21.76\% in \textbf{Multi-agent}, and the unluckiest seed (Worst) reaches 47.22\% and 47.78\%. Even \textsc{Consist-Strat} produces a median per-seed MaxErr of 14.42\% in \textbf{Multi-model} and 14.28\% in \textbf{Multi-agent}, with worst seeds reaching 34.61\% and 31.25\%. At 30\%, \textsc{Consist-Strat} still yields a worst seed error of 11.44\% and 10.5\%.

Two patterns explain why the worst-case error stays high. First, stratification controls the proportion of passes and fails in the subset, but does not control which instances are selected from each group. Random draws within a group can still skip behaviors the full test suite contains, so the worst-case error drops only 27--35\% from Random to \textsc{Consist-Strat} in \textbf{Multi-model} and \textbf{Multi-agent}.

Second, worst-case error does not scale with subset size. Going from 5\% to 30\% adds six times more instances but only cuts worst-case error by three times (from 34.61\% to 11.44\% in \textbf{Multi-model}).

In practice, looking only at the average error underestimates the risk of stochastic approaches. The averages in Table~\ref{tab:rq1-naive} summarize 500 random draws, but the spread of outcomes around each average is wide: at 10\% in \textbf{Multi-model}, \textsc{Consist-Strat} has a median RMSE of 0.05 but a typical-draw worst-case error of 9.63\% and a 95th-percentile worst-case error of 15.87\% (Table~\ref{tab:rq1-maxerr}), both far larger than the average suggests. A developer draws a subset once and gets a single outcome from this spread, which can be far worse than the average suggests.

In our use case, a developer changes their SWE-agent, for example by swapping the model or modifying the architecture, and evaluates the subset to decide whether to keep the change. One wrong estimate can trigger a rollback of a valid update or hide a real drop in resolve rate. A worst-seed error of 11\% at 30\% subset size leaves the result unreliable for the developers, motivating the deterministic methods in RQ2.

\subsubsection*{\textbf{Sensitivity Analysis}}
\label{sec:rq1_sensitivity}
\paragraph{By Temporal Window ($W$).}
The strongest baseline depends on how much historical information is available. At $W = 1$, where only a single past run is available, Consistency Stratification (\textsc{Consist-Strat}) cannot be applied, and Difficulty Stratification (\textsc{Diff-Strat}) instead serves as the strongest available baseline. When $W \geq 2$, \textsc{Consist-Strat} becomes the dominant baseline, using finer-grained historical pass counts to distinguish consistently brittle instances from consistently trivial ones.

\paragraph{Across Difficulty Levels.}
Outcome-aware stratification adapts to variation in test suite resolve rates. \textsc{Consist-Strat} samples proportionally from each pass-count group, so the subset preserves the difficulty mix regardless of the overall resolve rate. Random sampling can skew the subset toward any difficulty level, inflating MaxErr.

\paragraph{Across Subset Sizes and Datasets.}
The advantage of \textsc{Consist-Strat} over \textsc{Random} holds across all evaluated subset sizes, from 5\% to 30\%, and is statistically significant at every size ($p < 0.001$ after Holm correction, $\delta = 1.000$). The ranking \textsc{Consist-Strat} $>$ \textsc{Diff-Strat} $>$ \textsc{Random} is consistent across all datasets, suggesting that consistency-based stratification captures a general property of the test suite rather than an artifact of any particular temporal split.

\medskip
\noindent\fbox{%
\parbox{\dimexpr\linewidth-2\fboxsep-2\fboxrule}{%
\textbf{Summary of RQ1:} Consistency Stratification (\textsc{Consist-Strat}) is the strongest baseline, reducing median RMSE by 31--51\% relative to Random ($\delta = 1.000$, $p < 0.001$). Under Dem\v{s}ar's paired dominance ranking with Holm correction, \textsc{Consist-Strat} significantly outperforms all other baselines at every subset size. Repository Stratification provides negligible benefits ($<2\%$). All baseline approaches retain substantial worst-case risk. Even \textsc{Consist-Strat} yields a worst-case error of about 14\% on a typical draw at 5\% subset size, rising to 31--35\% on its worst seed, and its worst seed still reaches 10--11\% at a 30\% subset size.
}%
}

%% file: Sections/RQ2.tex
\subsection{RQ2: To what extent do trajectory-aware embedding methods improve subset selection compared to the strongest baseline?}
\label{sec:results_rq2}

\subsubsection*{\textbf{Motivation:}} 
As we saw in RQ1, at a 5\% subset size the best baseline produces a worst-case error of about 14\% on a typical draw and up to 35\% on its worst draw. RQ2 investigates whether we can reduce worst-case error by incorporating the agent's step-by-step behavior (i.e., its trajectory) from historical full runs rather than relying solely on its final pass/fail outcome. If the specific way an agent solves an instance provides a more reliable signal for subset selection, then our trajectory-aware methods should outperform the strongest approaches identified in RQ1.

\begin{table*}[t]
\centering
\caption{RQ2: RMSE of Embedding-Within-Strata methods vs.\ the strongest baseline (\textsc{Consist-Strat}), shown as a reference row. We show the pooled variant of each algorithm, which matches or outperforms the time-series variant. Single-setup results are consistent and omitted for space (see replication package). The best value per column, subset size, and dataset is \textbf{bolded}. Med = median RMSE. $\Delta\%_R$ = relative change in mean RMSE vs.\ \textsc{Consist-Strat} (negative = lower error). $p$ = Holm-adjusted Wilcoxon. $\delta$ = Cliff's delta~\cite{romano2006appropriate}. MaxErr is reported separately in Table~\ref{tab:rq2-maxerr}.}
\label{tab:rq2-embedding}
\resizebox{\textwidth}{!}{
\begin{tabular}{lrrrrrrrr}
\toprule
 & \multicolumn{4}{c}{\textbf{Multi-Model}} & \multicolumn{4}{c}{\textbf{Multi-Agent}} \\
\cmidrule(lr){2-5} \cmidrule(lr){6-9}
Method & Med & $\Delta\%_R$ & $p$ & $\delta$ & Med & $\Delta\%_R$ & $p$ & $\delta$ \\
\midrule
\multicolumn{9}{l}{\textbf{Subset size: 5\%}} \\
\midrule
\textsc{Consist-Strat} (ref.) & 0.0746 & -- & -- & -- & 0.0633 & -- & -- & -- \\
Centroid Pooled & 0.0679 & \textbf{$-$8.4} & $<$0.001 & +0.456 (M) & 0.0591 & $-$5.7 & $<$0.001 & +0.366 (M) \\
Core/Edge Pooled & \textbf{0.0677} & $-$7.4 & $<$0.001 & +0.372 (M) & \textbf{0.0586} & \textbf{$-$6.6} & $<$0.001 & +0.454 (M) \\
Medoid Pooled & 0.0709 & +1.4 & 0.76 & -- & 0.0624 & +0.3 & 1.00 & -- \\
FL Pooled & 0.0714 & +1.3 & 0.29 & -- & 0.0650 & +6.3 & $<$0.001 & $-$0.108 (N) \\
KS Pooled & 0.0716 & $-$2.3 & $<$0.001 & +0.208 (S) & 0.0605 & $-$2.8 & $<$0.001 & +0.282 (S) \\
\midrule
\multicolumn{9}{l}{\textbf{Subset size: 10\%}} \\
\midrule
\textsc{Consist-Strat} (ref.) & 0.0498 & -- & -- & -- & 0.0422 & -- & -- & -- \\
Centroid Pooled & \textbf{0.0431} & \textbf{$-$11.4} & $<$0.001 & +0.492 (L) & \textbf{0.0400} & \textbf{$-$3.1} & $<$0.001 & +0.222 (S) \\
Core/Edge Pooled & 0.0444 & $-$8.6 & $<$0.001 & +0.470 (M) & 0.0401 & $-$2.6 & $<$0.001 & +0.158 (S) \\
Medoid Pooled & 0.0465 & $-$3.0 & $<$0.001 & +0.234 (S) & 0.0407 & $-$0.2 & 0.78 & -- \\
FL Pooled & 0.0472 & +0.2 & 0.05 & -- & 0.0424 & +5.2 & $<$0.001 & $-$0.150 (S) \\
KS Pooled & 0.0498 & +1.4 & 1.00 & -- & 0.0427 & +3.1 & $<$0.001 & $-$0.160 (S) \\
\midrule
\multicolumn{9}{l}{\textbf{Subset size: 20\%}} \\
\midrule
\textsc{Consist-Strat} (ref.) & 0.0331 & -- & -- & -- & 0.0275 & -- & -- & -- \\
Centroid Pooled & 0.0313 & $-$2.8 & $<$0.001 & +0.220 (S) & 0.0279 & +1.6 & 0.39 & -- \\
Core/Edge Pooled & \textbf{0.0306} & \textbf{$-$5.4} & $<$0.001 & +0.298 (S) & 0.0287 & +6.9 & $<$0.001 & $-$0.350 (M) \\
Medoid Pooled & 0.0327 & +5.9 & $<$0.001 & $-$0.024 (N) & \textbf{0.0268} & \textbf{$-$1.4} & $<$0.001 & +0.186 (S) \\
FL Pooled & 0.0329 & +5.9 & $<$0.001 & $-$0.022 (N) & 0.0274 & +1.5 & 1.00 & -- \\
KS Pooled & 0.0330 & +1.3 & 0.84 & -- & 0.0282 & +3.3 & $<$0.001 & $-$0.106 (N) \\
\midrule
\multicolumn{9}{l}{\textbf{Subset size: 30\%}} \\
\midrule
\textsc{Consist-Strat} (ref.) & 0.0253 & -- & -- & -- & 0.0208 & -- & -- & -- \\
Centroid Pooled & \textbf{0.0225} & \textbf{$-$7.0} & $<$0.001 & +0.338 (M) & 0.0216 & +3.2 & $<$0.001 & $-$0.072 (N) \\
Core/Edge Pooled & 0.0234 & $-$4.6 & $<$0.001 & +0.304 (S) & 0.0228 & +12.8 & $<$0.001 & $-$0.548 (L) \\
Medoid Pooled & 0.0264 & +11.4 & $<$0.001 & $-$0.176 (S) & 0.0207 & +0.5 & 1.00 & -- \\
FL Pooled & 0.0261 & +10.8 & $<$0.001 & $-$0.138 (N) & 0.0208 & +1.7 & 1.00 & -- \\
KS Pooled & 0.0245 & +0.2 & 1.00 & -- & \textbf{0.0206} & \textbf{$-$0.3} & 0.26 & -- \\
\bottomrule
\end{tabular}}
\end{table*}

\begin{table*}[t]
\centering
\caption{RQ2: MaxErr (\%) of Embedding-Within-Strata methods vs.\ the per-seed MaxErr distribution of \textsc{Consist-Strat} (Table~\ref{tab:rq1-maxerr}). Each embedding method produces a single deterministic MaxErr. $\Delta\%_{Med}$ / $\Delta\%_{P95}$ / $\Delta\%_{W}$ = relative change vs.\ \textsc{Consist-Strat}'s median / 95th-percentile / worst per-seed MaxErr (negative = lower error). The best value per column, subset size, and dataset is \textbf{bolded}.}
\label{tab:rq2-maxerr}
\resizebox{\textwidth}{!}{
\begin{tabular}{lrrrrrrrr}
\toprule
 & \multicolumn{4}{c}{\textbf{Multi-Model}} & \multicolumn{4}{c}{\textbf{Multi-Agent}} \\
\cmidrule(lr){2-5} \cmidrule(lr){6-9}
Method & MaxErr & $\Delta\%_{Med}$ & $\Delta\%_{P95}$ & $\Delta\%_{W}$ & MaxErr & $\Delta\%_{Med}$ & $\Delta\%_{P95}$ & $\Delta\%_{W}$ \\
\midrule
\multicolumn{9}{l}{\textbf{Subset size: 5\%}} \\
\midrule
\textsc{Centroid Pooled} & \textbf{12.92} & \textbf{$-$10.4} & \textbf{$-$45.8} & \textbf{$-$62.7} & 13.75 & $-$3.7 & $-$37.9 & $-$56.0 \\
\textsc{Core/Edge Pooled} & 12.92 & $-$10.4 & $-$45.8 & $-$62.7 & \textbf{13.34} & \textbf{$-$6.6} & \textbf{$-$39.8} & \textbf{$-$57.3} \\
\textsc{Medoid Pooled} & 13.44 & $-$6.8 & $-$43.6 & $-$61.2 & 13.75 & $-$3.7 & $-$37.9 & $-$56.0 \\
\textsc{FL Pooled} & 13.57 & $-$5.9 & $-$43.1 & $-$60.8 & 14.06 & $-$1.5 & $-$36.5 & $-$55.0 \\
\textsc{KS Pooled} & 14.20 & $-$1.5 & $-$40.4 & $-$59.0 & 13.92 & $-$2.5 & $-$37.1 & $-$55.4 \\
\midrule
\multicolumn{9}{l}{\textbf{Subset size: 10\%}} \\
\midrule
\textsc{Centroid Pooled} & \textbf{8.56} & \textbf{$-$11.1} & \textbf{$-$46.1} & \textbf{$-$62.7} & 8.95 & $-$6.8 & $-$40.1 & $-$57.9 \\
\textsc{Core/Edge Pooled} & 8.92 & $-$7.4 & $-$43.8 & $-$61.1 & 9.05 & $-$5.7 & $-$39.4 & $-$57.4 \\
\textsc{Medoid Pooled} & 9.09 & $-$5.6 & $-$42.7 & $-$60.3 & \textbf{8.91} & \textbf{$-$7.2} & \textbf{$-$40.4} & \textbf{$-$58.0} \\
\textsc{FL Pooled} & 9.22 & $-$4.3 & $-$41.9 & $-$59.8 & 9.24 & $-$3.7 & $-$38.1 & $-$56.5 \\
\textsc{KS Pooled} & 9.82 & $+$2.0 & $-$38.1 & $-$57.2 & 9.40 & $-$2.1 & $-$37.1 & $-$55.7 \\
\midrule
\multicolumn{9}{l}{\textbf{Subset size: 20\%}} \\
\midrule
\textsc{Centroid Pooled} & \textbf{5.90} & \textbf{$-$8.0} & \textbf{$-$44.0} & \textbf{$-$61.0} & 6.25 & $-$0.7 & $-$36.1 & $-$54.8 \\
\textsc{Core/Edge Pooled} & 6.02 & $-$6.1 & $-$42.8 & $-$60.2 & 6.37 & $+$1.1 & $-$34.9 & $-$54.0 \\
\textsc{Medoid Pooled} & 6.40 & $-$0.1 & $-$39.2 & $-$57.6 & \textbf{5.69} & \textbf{$-$9.7} & \textbf{$-$41.9} & \textbf{$-$58.9} \\
\textsc{FL Pooled} & 6.40 & $-$0.1 & $-$39.2 & $-$57.6 & 5.79 & $-$8.1 & $-$40.8 & $-$58.2 \\
\textsc{KS Pooled} & 6.40 & $-$0.1 & $-$39.2 & $-$57.6 & 5.88 & $-$6.6 & $-$39.9 & $-$57.5 \\
\midrule
\multicolumn{9}{l}{\textbf{Subset size: 30\%}} \\
\midrule
\textsc{Centroid Pooled} & \textbf{4.43} & \textbf{$-$9.3} & \textbf{$-$44.7} & \textbf{$-$61.2} & 4.86 & $+$1.8 & $-$34.6 & $-$53.7 \\
\textsc{Core/Edge Pooled} & 4.73 & $-$3.2 & $-$41.0 & $-$58.6 & 5.05 & $+$5.8 & $-$32.0 & $-$51.9 \\
\textsc{Medoid Pooled} & 5.11 & $+$4.6 & $-$36.2 & $-$55.3 & 4.32 & $-$9.5 & $-$41.8 & $-$58.8 \\
\textsc{FL Pooled} & 5.09 & $+$4.2 & $-$36.5 & $-$55.5 & 4.35 & $-$9.0 & $-$41.5 & $-$58.6 \\
\textsc{KS Pooled} & 4.93 & $+$0.9 & $-$38.5 & $-$56.9 & \textbf{4.30} & \textbf{$-$9.9} & \textbf{$-$42.1} & \textbf{$-$59.0} \\
\bottomrule
\end{tabular}}
\end{table*}

\subsubsection*{\textbf{Approach:}} 

We replace the random sampling step from RQ1 with a targeted selection process based on the agent's trajectory, then compare the trajectory-aware approach directly against the strongest baseline from RQ1 (\textsc{Consist-Strat}). We use the same evaluation setup (all three datasets, four subset sizes, and temporal cross-validation) as for RQ1. Because our geometric methods and \textsc{Consist-Strat} group instances by their past resolve counts in the same way, any change in average accuracy (RMSE) or worst-case risk (MaxErr) can be directly attributed to the selection method.

Instead of randomly drawing from the test outcome groups, we apply the five geometric algorithms of Section~\ref{subsubsec:geometric_selection} to the trajectory embeddings, using either the pooled or the time-series variant (Section~\ref{subsec:approach_vectorization}), yielding ten configurations in total.

We compare all ten configurations against \textsc{Consist-Strat} using the pairwise Wilcoxon test, Holm correction, and Cliff's $\delta$ (Section~\ref{subsec:statistical_analysis}).

\subsubsection*{\textbf{Finding 2.1:}} \textbf{Trajectory-aware embedding methods reduce mean RMSE by up to 11.4\% over the strongest baseline.}
\label{sec:rq2_finding1}
Trajectory-aware methods that apply geometric selection within historical test outcome groups (\textit{Embedding-Within-Strata}, Section~\ref{subsubsec:geometric_selection}) significantly outperform the strongest baseline (i.e., \textsc{Consist-Strat}) at small subset sizes (5\% and 10\%) in both datasets ($p < 0.001$, Wilcoxon with Holm correction, Table~\ref{tab:rq2-embedding}). At larger subset sizes (20\%--30\%), the advantage holds in \textbf{Multi-model} but disappears in \textbf{Multi-agent}, where \textsc{Centroid Pooled} is on par with \textsc{Consist-Strat} (negligible $\delta$) and \textsc{Core/Edge Pooled} falls behind it. Table~\ref{tab:rq2-embedding} shows that \textsc{Centroid Pooled} and \textsc{Core/Edge Pooled} consistently rank among the best-performing methods at 5\% and 10\%.

\textsc{Centroid Pooled} uses the same historical pass-count groups as \textsc{Consist-Strat}, but instead of sampling randomly within each group, it selects instances closest to the group's embedding center.

At a 10\% subset size in the \textbf{Multi-model} dataset, Table~\ref{tab:rq2-embedding} shows that \textsc{Centroid Pooled} reduces mean RMSE by 11.4\% relative to \textsc{Consist-Strat} ($p < 0.001$, $\delta = 0.492$, large effect size). Figure~\ref{fig:rq2-compare} (top row) compares \textsc{Centroid Pooled} against \textsc{Consist-Strat} on each of the 1,000 synthetic test suite distributions generated (Section~\ref{subsec:synthetic_distribution}), each a set of 250 benchmark instances. At 5\% and 10\% subset sizes, \textsc{Centroid Pooled} produces lower RMSE than \textsc{Consist-Strat} on the large majority of distributions, confirming that the improvement holds across individual distributions, not only in aggregate.

\subsubsection*{\textbf{Finding 2.2:}} \textbf{Trajectory-aware methods reduce MaxErr by 4--11\% relative to the typical draw and 38--46\% relative to the 95th-percentile draw, reducing the risk of large errors in baseline approaches.}
\label{sec:rq2_finding2}
Although the gains in average accuracy are meaningful, the trajectory-aware embedding methods also reduce the worst-case error.

As shown in RQ1, baseline approaches exhibit high MaxErr because random sampling can omit important behavioral profiles. Trajectory-aware methods avoid the high MaxErr
of random sampling by selecting instances deterministically from the structure of the embedding space rather than relying on random draws.

\begin{figure*}[t]
\centering
\includegraphics[width=0.8\textwidth]{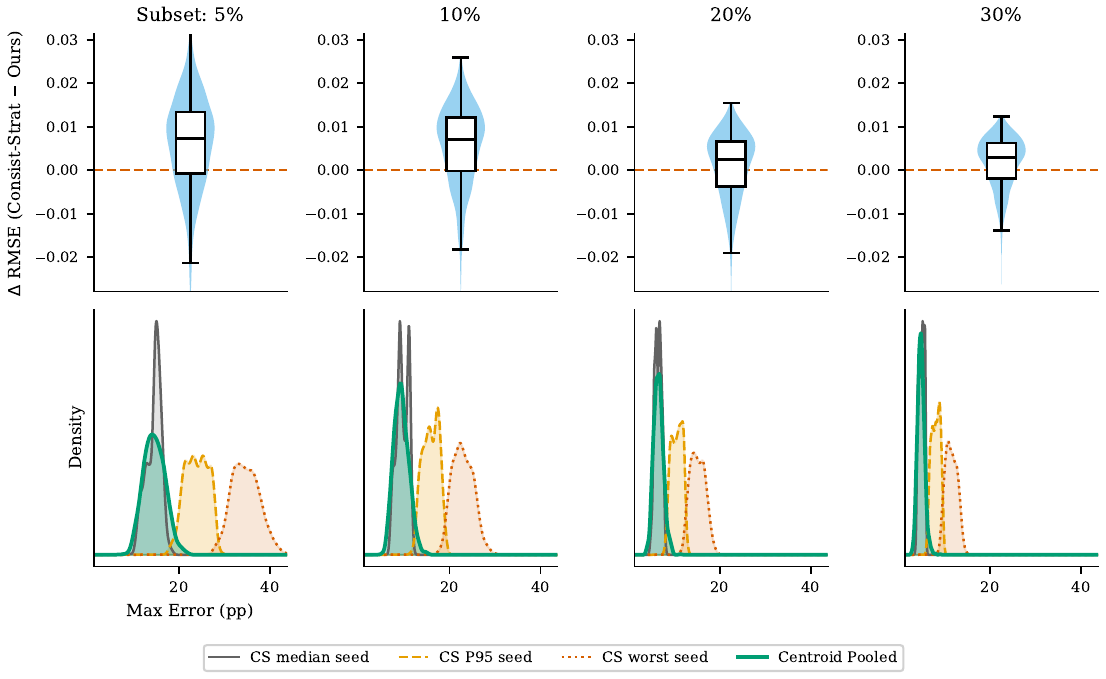}
\caption{Performance of the strongest trajectory-aware embedding method (\textsc{Centroid Pooled}) vs.\ the strongest baseline (\textsc{Consist-Strat}) across 1,000 synthetic evaluations (Multi-model dataset). \textbf{Top Row:} The paired difference in RMSE ($\Delta$ RMSE = Baseline $-$ Ours). The majority of the interquartile range sits above the break-even line ($y=0$), showing that \textsc{Centroid Pooled} produces lower RMSE than the baseline on the majority of individual distributions ($p < 0.001$ across all subset sizes, Table~\ref{tab:rq2-embedding}). \textbf{Bottom Row:} Per-seed MaxErr density for \textsc{Consist-Strat} at three aggregation levels (median, P95, and worst across 500 seeds) compared to the deterministic MaxErr of \textsc{Centroid Pooled}. The deterministic method's distribution sits below the baseline's median-seed density, eliminating the long tail of high-error seeds entirely.}
\label{fig:rq2-compare}
\end{figure*}

Table~\ref{tab:rq2-maxerr} shows that \textsc{Centroid Pooled} reduces MaxErr by 4--11\% relative to the typical draw and 38--46\% relative to the 95th-percentile draw, across the \textbf{Multi-model} and \textbf{Multi-agent} datasets at the 5\% and 10\% subset sizes. Figure~\ref{fig:rq2-compare} illustrates the MaxErr reduction clearly. At a 10\% subset size, the strongest baseline's typical draw produces worst-case errors of about 10 percentage points, with the 95th-percentile draw reaching 15-16 percentage points, whereas \textsc{Centroid Pooled} limits MaxErr to roughly 9\%. 

To understand the MaxErr reduction, we examine a worst-case draw from \textsc{Consist-Strat} at 5\% subset size. \textsc{Consist-Strat} drew 12 instances where every instance
passed on the Kimi K2 run, producing a subset resolve rate of 1.00 against a test suite rate of 0.43 on Kimi K2, an error off by 57\%. \textsc{Centroid Pooled}, applied to the same test outcome groups, selected instances closest to the embedding center of each group, producing a Kimi K2 subset resolve rate of 0.42 and an error of 1\%. \textsc{Consist-Strat} can draw an unrepresentative subset by chance, whereas \textsc{Centroid Pooled} selects instances close to the center of each test outcome group, keeping the subset resolve rate close to the test suite resolve rate. This worst-case benefit extends even to the geometric algorithms that do not improve average accuracy: at a 5\% subset size in \textbf{Multi-model}, Medoid Pooled and FL Pooled perform on par with \textsc{Consist-Strat} on RMSE (+1.4\% and +1.3\%, not statistically significant).

By deterministically selecting the geometric center, trajectory-aware selection reduces the chance that an important behavioral profile is entirely excluded from the subset.

\subsubsection*{\textbf{Sensitivity Analysis}}
\label{sec:rq2_sensitivity}

\paragraph{By Temporal Window ($W$).}
The advantage of embedding-based selection over the baseline holds regardless of how much historical data is available. With only one prior run ($W = 1$), instances can only be grouped into pass or fail. With two or more prior runs ($W \geq 2$), instances can be grouped by their exact number of passes. In both cases, \textsc{Centroid Pooled} produces lower RMSE than the strongest baseline available at that window size, \textsc{Diff-Strat} at $W = 1$ and \textsc{Consist-Strat} at $W \geq 2$, indicating that trajectory embeddings capture behavioral information beyond what historical resolve rates alone provide.

\paragraph{Across Difficulty Levels.}
The strongest trajectory-aware method varies with the test suite resolve rate. Across the nine difficulty levels evaluated in Section~\ref{subsec:synthetic_distribution}, \textsc{Centroid Pooled} achieves the lowest RMSE when the test suite difficulty level falls in the middle range (40\% -- 60\% source resolve rate), where the test suite contains a balanced mix of passing and failing instances.  \textsc{Core/Edge Pooled} achieves the lowest RMSE more often at the lowest difficulty levels (10\% -- 30\%  source resolve rate), where the test suite is dominated by failing instances and selection benefits from broader coverage of the embedding space. In practice, \textsc{Centroid Pooled} is the strongest general-purpose default, whereas test suites skewed toward failure may benefit more from core-edge selection.

\paragraph{Across Subset Sizes and Datasets.}
Embedding-based methods provide the greatest improvement over \textsc{Consist-Strat} at subset sizes of 5\% and 10\%, where the small subset sizes leave random sampling with too few instances to reliably represent the full test suite. As the subset size increases to 20\% and 30\%, random sampling includes a larger fraction of the test suite by chance, narrowing the RMSE gap between approaches (Table~\ref{tab:rq2-embedding}). At 30\%
subset size, baseline methods occasionally match or slightly exceed the average accuracy of embedding methods (for example, \textsc{Consist-Strat} reaches a median RMSE of 0.0208 in
\textbf{Multi-agent}, compared to 0.0216 for \textsc{Centroid Pooled}). At 30\% subset size, the MaxErr advantage narrows: Centroid Pooled reduces MaxErr by 9\% in \textbf{Multi-model} but shows no improvement in \textbf{Multi-agent} relative to the typical Consist-Strat draw.

\medskip
\noindent\fbox{%
\parbox{\dimexpr\linewidth-2\fboxsep-2\fboxrule}{%
\textbf{Summary of RQ2:} \textsc{Centroid Pooled}, applied
within test outcome groups, reduces mean RMSE by up to 11.4\%
($\delta = 0.492$, $p < 0.001$) and MaxErr by 4--11\% relative to the typical draw and 38--46\% relative to the 95th-percentile draw of \textsc{Consist-Strat} at 5\% and 10\% subset size.
}%
}

%% file: Sections/RQ3.tex
\subsection{RQ3: What drives the improvement of trajectory-aware methods over the baselines?}
\label{sec:results_rq3}

\subsubsection*{\textbf{Motivation:}} RQ3 is an ablation study that isolates which component of the trajectory-aware method produces the gains observed in RQ2. The improvement in RMSE and MaxErr could come from trajectory embeddings, historical outcome stratification, the fact that geometric selection produces a fixed subset, or a combination of these.

\begin{figure*}[!t]
    \centering
    \includegraphics[width=\linewidth]{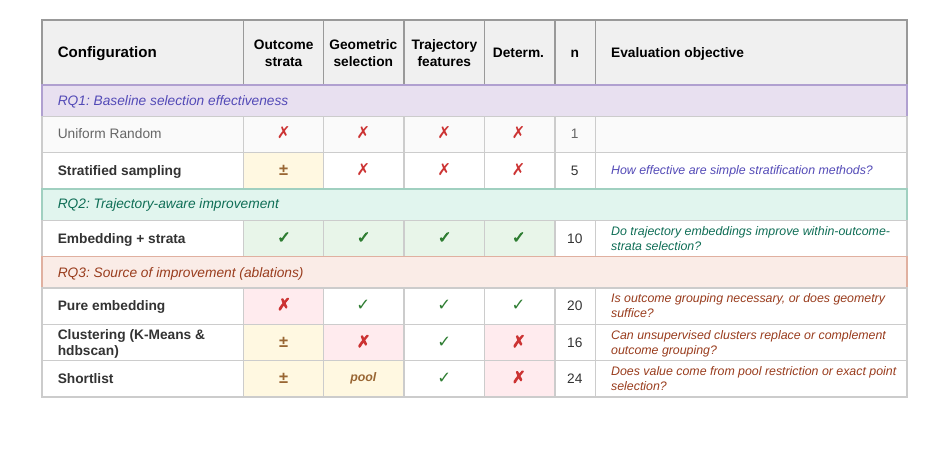}
    \caption{Ablation matrix of subset selection configurations. 
    Three binary design decisions define the method space. 
    RQ1 establishes baseline selection effectiveness, RQ2 evaluates 
    trajectory-aware improvement, and RQ3 ablations each 
    modify one or two toggles from the full configuration 
    to isolate individual component contributions. 
    $n$ = number of configurations per family, 76 in total. Uniform Random is a single method, Stratified sampling covers the five baselines that group instances before sampling, and Embedding + strata is 5 geometric algorithms $\times$ 2 embedding variants. Section~\ref{sec:results_rq3} derives the three ablation counts. Every configuration is evaluated at all four subset sizes.}
    \label{fig:ablation_matrix}
\end{figure*}

\subsubsection*{\textbf{Approach:}} 

To identify which components of the pipeline drive the gains from trajectory-aware selection, we conduct a systematic ablation study. Figure~\ref{fig:ablation_matrix} summarizes the ablation matrix: each row defines a configuration family, and
column $n$ reports the number of configurations evaluated within each family, totaling 76 configurations across all families. For each ablation, we compare the best configuration of the family against \textsc{Centroid Pooled} on the same 1,000 distributions with the pairwise Wilcoxon test, Holm correction, and Cliff's $\delta$ (Section~\ref{subsec:statistical_analysis}).
We test three specific design decisions, followed by a fourth check on determinism:

\begin{enumerate}
\item \textbf{Removing Outcome Groups (Pure Embedding):} We test if grouping instances by past success rates is necessary. For this ablation, we apply our geometric selection algorithms directly to the full test suite without any prior historical grouping to test the value of geometric selection alone. We evaluate each algorithm with and without a post-hoc calibration step that adjusts the selected subset so its pass-count composition matches the full test suite, giving 5 algorithms $\times$ 2 embedding variants $\times$ 2 calibration settings = 20 configurations.
\item \textbf{Replacing Outcome Groups with Unsupervised Clustering:} We test if unsupervised clustering over trajectory embeddings can
replace outcome-based grouping. Instead of grouping instances by
their historical pass counts, we group them by behavioral
similarity using two clustering methods:
\begin{itemize}
    \item \textbf{K-Means~\cite{arthur2007kmeans}:} Partitions instances into $K$ groups by repeatedly assigning each instance to the nearest group center, then recomputing each center as the average of its assigned members. Because $K$
    must be specified in advance, we test values from 2 to 15 and select the $K$ that produces the most internally coherent groups, measured by the silhouette score~\cite{kaufman1990finding}.
    \item \textbf{HDBSCAN~\cite{mcinnes2017hdbscan}:} Finds groups by identifying regions where instances are densely packed together, without requiring the number of groups to be specified in advance. Instances in sparse regions are
    labeled as noise. To ensure all instances remain eligible for selection, we reassign noise instances to the nearest group center.
\end{itemize}

We run each clustering method over four feature spaces: the pooled and time-series trajectory embeddings, all ten trajectory features (e.g., number of steps, edits, errors, and tests), and a subset of five of those features. We include the subset because clustering on all ten features produces poorly separated groups. We rank the features by silhouette score, a measure of how well separated the resulting clusters are, and keep the five that score highest: the number of test steps, edit steps, errors, total steps, and unique files touched. We evaluate each combination twice, once clustering the full test suite and once clustering within each outcome group, giving 2 clustering methods $\times$ 4 feature spaces $\times$ 2 grouping settings = 16 configurations.

\item \textbf{Shortlisting vs. Exact Selection:} Our geometric methods directly select the instances for the final subset. To test whether the value of embeddings lies in selecting those exact instances or in narrowing the pool before selecting the final instances using another sampling method, we apply a two-step approach: first, use a geometric algorithm to select a larger candidate pool (either 1.5 or 2 times the target subset size), then randomly sample from that pool to fill the final subset. We test three geometric algorithms (Facility Location, Centroid, and Kennard-Stone), which are the three base algorithms from which the
remaining two (Medoid and Core/Edge) are derived, across both embedding types (pooled and time-series), with and without historical outcome grouping, at both pool multipliers (1.5$\times$ and 2$\times$), producing
$3 \times 2 \times 2 \times 2 = 24$ configurations, each
evaluated at all four subset sizes.
\end{enumerate}

\paragraph{Determinism Validation.}\label{sec:rq3_determinism} All geometric configurations are deterministic: given the same prior runs, embeddings, and subset size, they always select the same subset (Section~\ref{subsubsec:geometric_selection}). \textsc{Consist-Strat}, in contrast, draws a different random subset on every run. We validate using \textsc{Centroid Pooled}, the strongest geometric configuration in RQ2.

A fixed subset could produce lower RMSE than a random one simply by avoiding sampling variance, even without the trajectory embeddings. To check whether the embedding signal adds value beyond determinism, we run two additional tests.

First, we compare \textsc{Centroid Pooled} against 500 random draws of \textsc{Consist-Strat} with different random seeds, producing 500 RMSE values for each distribution. We then check where \textsc{Centroid Pooled}'s RMSE falls relative to the middle 95\% of those 500 values. Falling inside the range means
the two approaches are not distinguishable on that distribution. Falling below it means the deterministic subset is more accurate than nearly every random draw, and falling above it means the opposite. A fixed subset carrying no information about agent behavior would fall below the range on 2.5\% of distributions and above it on 2.5\%, so an asymmetry between the two indicates
a real difference. This test uses the runs where \textsc{Consist-Strat} applies, which are those with at least two prior runs.

Second, we test whether a fixed selection rule without trajectory embeddings reproduces the gains. \textit{Stability-Stratified} is a control method we construct for this test only, not one of the 76 configurations. It uses the same binary pass/fail groups as \textsc{Diff-Strat} and, within each group, selects the instances whose pass/fail outcome varied least across the training runs, so the pick is fixed and based on past outcomes but ignores the trajectories. We compare it against \textsc{Diff-Strat}, which draws at random from the identical groups, so the only difference between the two is the pick within each group. We use the pairwise Wilcoxon test and Cliff's $\delta$ for RMSE, and the per-seed MaxErr statistics of Section~\ref{subsec:metrics} for worst-case error. The test uses runs with at least two prior runs, since with a single prior run every instance has the same outcome variance and the rule cannot rank them.

\begin{table*}[t]
\centering
\caption{RQ3 Ablation: best configuration of each ablation family
compared against the full method (\textsc{Centroid Pooled [ES]})
at 10\% subset size. The families correspond to the three RQ3 ablation families in Figure~\ref{fig:ablation_matrix}. Suffixes mark the family: [ES] Embedding-Within-Strata, [PE] Pure Embedding, [CL] Clustering, [SL] Shortlist. Med = median RMSE across distributions. $\Delta$\%$_R$ = relative change in the ablation's mean RMSE against the full method (positive = the ablation has the higher error). Med reports medians while $\Delta$\% compares means, so the two are not derivable from each other. $p$ = Holm-adjusted Wilcoxon. $\delta$ = Cliff's delta.
\textbf{Multi-model} and \textbf{Multi-agent} datasets shown ($N{=}1{,}000$
distributions each).}
\label{tab:rq3-ablation}
\resizebox{\textwidth}{!}{
\begin{tabular}{llrrrrrrrr}
\toprule
Family & Best Config. & \multicolumn{4}{c}{Multi-model} & \multicolumn{4}{c}{Multi-agent} \\
\cmidrule(lr){3-6} \cmidrule(lr){7-10}
& & Med & $\Delta$\%$_R$ & $p$ & $\delta$ & Med & $\Delta$\%$_R$ & $p$ & $\delta$ \\
\midrule
Embedding-Within-Strata (full method) & \textbf{Centroid Pooled [ES]} & \textbf{0.0431} & --- & --- & --- & \textbf{0.0400} & --- & --- & --- \\
Pure Embedding & Medoid TS [PE] & 0.0634 & +36.7 & $<$0.001 & +0.588 (L) & 0.0573 & +30.5 & $<$0.001 & +0.778 (L) \\
Clustering & HDBSCAN Pooled [CL] & 0.0526 & +15.7 & $<$0.001 & +0.620 (L) & 0.0448 & +9.0 & $<$0.001 & +0.500 (L) \\
Shortlist & Centroid Pooled +Strata 1.5$\times$ [SL] & 0.0565 & +20.8 & $<$0.001 & +0.866 (L) & 0.0547 & +25.5 & $<$0.001 & +0.872 (L) \\
\bottomrule
\end{tabular}}
\end{table*}
\subsubsection*{\textbf{Finding 3.1:}} \textbf{Outcome grouping
is necessary, as removing it increases RMSE by 30--37\%.} \label{sec:rq3_finding1} Removing outcome grouping (the Pure Embedding ablation in
Table~\ref{tab:rq3-ablation} and Figure~\ref{fig:ablation_matrix})
increases RMSE by 36.7\% in the \textbf{Multi-model} dataset and 30.5\% in the \textbf{Multi-agent} dataset relative to \textsc{Centroid Pooled} (Table~\ref{tab:rq3-ablation}). Both differences are significant ($p < 0.001$, Holm-adjusted Wilcoxon) with large effect sizes ($\delta = 0.588$ and $0.778$). Without outcome grouping, geometric selection picks the most behaviorally typical
instances from the full test suite. When the test suite contains many easy instances, the typical instances are easy. When the test suite contains many hard instances, the typical instances are hard. The resulting subset over-represents whichever difficulty level dominates the test suite, biasing the resolve-rate estimate. The outcome grouping first fixes the difficulty composition, then geometric selection picks typical
behavior within each difficulty level. 

Outcome grouping and embedding geometry address different problems: outcome grouping controls the difficulty composition of the subset, and embedding geometry controls behavioral representativeness within each difficulty level.

\subsubsection*{\textbf{Finding 3.2:}} \textbf{Behavioral clustering cannot replace outcome grouping.}
\label{sec:rq3_finding2}
The best clustering configuration (\textsc{HDBSCAN Pooled}) has
15.7\% higher RMSE in \textbf{Multi-model} and 9.0\% higher RMSE
in \textbf{Multi-agent} than \textsc{Centroid Pooled} (Table~\ref{tab:rq3-ablation}). Both differences are significant ($p < 0.001$) with large effect sizes ($\delta = 0.620$ and $0.500$).
Clustering groups' membership is based on similarity in past behavior, whether measured by trajectory embeddings or by trajectory features such as the number of steps, edits, errors, and tests. Instances that behave similarly can still resolve at different rates, so a behavioral cluster mixes easy and hard instances and the subset no longer preserves the test suite's difficulty composition. Historical pass counts directly measure the resolve frequency, which is what the subset must preserve to estimate the full test suite's resolve rate.

\subsubsection*{\textbf{Finding 3.3:}} \textbf{Exact geometric selection outperforms
random sampling from the narrowed pool.}
\label{sec:rq3_finding3}
The best shortlist configuration has 20.8\% higher RMSE in
\textbf{Multi-model} and 25.5\% higher RMSE in \textbf{Multi-agent}
than \textsc{Centroid Pooled} (Table~\ref{tab:rq3-ablation}). Both differences are significant ($p < 0.001$) with large effect sizes ($\delta = 0.866$ and $0.872$). Geometric selection
narrows the pool to behaviorally representative candidates, but
the random sampling step that follows can still draw an
unrepresentative subset from within the pool. Random sampling is effective when instances are exchangeable
within a group (as in \textsc{Consist-Strat}). Once geometric
selection has identified specific representative instances,
replacing them with random draws discards the selection signal
and reintroduces sampling variance.

\subsubsection*{\textbf{Finding 3.4:}} \textbf{Trajectory embeddings improve accuracy beyond what determinism alone provides.}
\label{sec:rq3_finding4}
The two tests of the Determinism Validation approach separate the effect of determinism from the effect of the trajectory embeddings.

In the first test, \textsc{Centroid Pooled} falls below the middle 95\% of the \textsc{Consist-Strat} draws on 20.4\% of distributions in \textbf{Multi-model} and 30.7\% in \textbf{Multi-agent} at a 10\% subset size, and above it on 1.9\%
and 0.6\%. At a 5\% subset size the figures are 18.2\% and 26.7\% below against 0.3\% and 0.5\% above. A fixed subset with no behavioral signal would land outside the range equally often in both directions, so the asymmetry shows that the trajectory embeddings, and not the fixed subset alone, drive the result. 

In the second test, the fixed outcome-based selection is not a substitute for the embedding selection. At 5\% and 10\% subset size, \textit{Stability-Stratified}'s mean RMSE ranges from 20.5\% lower to 18.3\% higher than \textsc{Diff-Strat}'s across the two datasets, and its worst-case error from 20.3\% lower to 11.4\% higher than a typical \textsc{Diff-Strat} draw. At 20\% and 30\%, it is worse than the random draw in both datasets, by 19--90\% on mean RMSE and 21--69\% on typical worst-case error ($p < 0.001$, $|\delta|$ from 0.37 to 0.72). \textsc{Centroid Pooled} lowers both RMSE and worst-case error relative to \textsc{Consist-Strat} at 5\% and 10\% in both datasets (Tables~\ref{tab:rq2-embedding} and~\ref{tab:rq2-maxerr}) and stays on par at larger sizes. A fixed subset is therefore not what produces the gain. The selection within each outcome group is.

\medskip
\noindent\fbox{%
\parbox{\dimexpr\linewidth-2\fboxsep-2\fboxrule}{%
\textbf{Summary of RQ3:} Trajectory-aware subset selection degrades under all three ablations relative to \textsc{Centroid Pooled}. Removing outcome grouping (Pure Embedding) raises RMSE by 30--37\%. Replacing outcome grouping with unsupervised clustering raises RMSE by 9--16\%. Narrowing the pool geometrically and then sampling randomly from it raises RMSE by 21--26\%. Determinism is not the source of the gain: a fixed selection based on past outcomes alone is not consistently better than random sampling and falls behind it from 20\% subset size up, while the embedding-based selection improves on random sampling at 5\% and 10\% subset sizes}%
}

%% file: Sections/RQ4.tex
\subsection{RQ4: How much does subset selection
reduce evaluation cost?}
\label{sec:results_rq4}

\noindent\textbf{Motivation:} RQ1--RQ3 show that
trajectory-aware subset selection reduces estimation
error at small subset sizes. Instances differ widely in how many tokens they consume, so a 10\% subset does not automatically cost 10\% of the full benchmark. A 10\%
subset that happens to contain the most expensive
trajectories saves far less than 90\%, and a 10\%
subset that picks the cheapest trajectories saves
more than 90\% by under-sampling the harder instances.
RQ4 measures to what extent subset cost scales proportionally
with subset size, so the accuracy gains observed in
Sections~\ref{sec:results_rq1}--\ref{sec:results_rq3}
translate into proportional cost savings.

\noindent\textbf{Approach:} To quantify cost savings,
we compute the cumulative token cost (input and output) of
each trajectory across all six runs of the
\textbf{Multi-model} dataset. We use the \textbf{Multi-model} dataset because it matches the development scenario RQ4 targets, where the framework stays fixed while the model or its settings change. The \textbf{Single-setup} dataset reruns one configuration, so there is no cost change to measure, and the \textbf{Multi-agent} dataset changes the framework, so cost differences reflect the framework rather than the instances. Each trajectory produces a single token count: the total count of all input and output tokens across every step.
The per-trajectory total reflects the real API cost: each step
re-sends the full conversation history as input, so we count every message once for each later step that includes it in its context. Tokens therefore
accumulate and the context grows quickly across
steps~\cite{gao2025morewithless, agentdiet}. Following
standard agent framework
behavior~\cite{lindenbauer2025complexitytraps}, we truncate any single tool output longer than 10{,}000 tokens before adding it to the conversation history.

To measure how cost savings scale with subset size, we draw 10{,}000 random subsets of instances per run at each subset size ($k = 5, 10, 20, 30$) and measure each subset's token cost as a share of the run's total token cost across all 500 instances. Averaging these 10{,}000 draws gives the cost share a subset of that size consumes by chance. We then compare the cost share of the \textsc{Centroid Pooled} 10\% subset against the 10{,}000 random draws at the same size. The random draws cover all four subset sizes (Table~\ref{tab:cost_savings}), while we run the \textsc{Centroid Pooled} comparison at 10\% only, the subset size we use as the headline setting throughout the paper (Table~\ref{tab:rq3-ablation} and Section~\ref{sec:conclusion}).

Table~\ref{tab:agent_cost} summarizes the cost profile of the six \textbf{Multi-model} runs. Running all 500 SWE-Bench Verified instances once costs between 335M and 945M tokens depending on the run, and 3.44 billion tokens across all six runs. Input tokens account for 99.3\% of the total volume. Cost scales disproportionately with agent steps: Claude~4~Sonnet takes 2.2$\times$ as many steps as GPT-5 (69.7 vs.\ 31.2) but costs 2.7$\times$ more per instance (1.89M vs.\ 0.70M), consistent with the quadratic accumulation described in prior work~\cite{gao2025morewithless, agentdiet}.

\begin{table}[t]
\centering
\caption{Per-run cost profile across 500 SWE-Bench
Verified instances. All runs use the OpenHands
framework with different models. Token counts reflect
cumulative cost (input re-transmission at
every turn). Total (M) = total \#tokens in millions across all 500
instances. Mean (M) = mean \#tokens per instance in
millions. Steps = mean number of agent steps per
instance. I/O Ratio = ratio of \#input tokens to \#output
tokens.}
\label{tab:agent_cost}
\small
\begin{tabular}{lrrrr}
\toprule
\textbf{OpenHands Run} & \textbf{Total (M)} & \textbf{Mean (M)} & \textbf{Steps} & \textbf{I/O Ratio} \\
\midrule
Sonnet 3.5 (Oct '24)        & 335  & 0.67 & 26.2 & 128$\times$ \\
Sonnet 3.7 (Apr '25)       & 369  & 0.74 & 37.2 & 109$\times$ \\
Devstral (May '25)      & 758  & 1.52 & 49.3 & 166$\times$ \\
Sonnet 4 (May '25)      & 945  & 1.89 & 69.7 & 147$\times$ \\
Kimi K2 (Jul '25)       & 684  & 1.37 & 55.9 & 114$\times$ \\
GPT-5 (Aug '25)         & 351  & 0.70 & 31.2 & 234$\times$ \\
\midrule
\textbf{Aggregate}      & \textbf{3{,}442} & \textbf{1.15} & \textbf{44.9} & \textbf{140$\times$} \\
\bottomrule
\end{tabular}
\end{table}

\noindent\textbf{Finding 4.1: Subset cost is proportional to subset size, i.e., a $k$\% subset consumes $k$\% of the full benchmark's token cost on average.}
The result holds across all runs and subset sizes
(Table~\ref{tab:cost_savings}). The reported range is the spread across the 10{,}000 draws within a run. Individual instances differ in cost by up to 492$\times$, so one draw can land well above or below its expected $k$\% share, while the average across the 10{,}000 draws stays at $k$\%.

\begin{table}[t]
\centering
\caption{Empirical cost savings by subset size. For each of the
six runs, we draw 10{,}000 random subsets of size $k$\% and
compute each subset's share of the run's total token cost.
Cost Share is the mean over the 10{,}000 draws, averaged across the six runs, and it matches the subset size to one decimal place. 95\% Range is the middle 95\% of the draws (2.5th to 97.5th percentile), taken from the run with the widest spread, and it shows how far a single draw can fall from the mean. Savings = $100 - \text{Cost Share}$.}
\label{tab:cost_savings}
\small
\begin{tabular}{rrrr}
\toprule
\textbf{Subset} & \textbf{Cost Share ($\mu$)} & \textbf{95\% Range} & \textbf{Savings} \\
\midrule
5\%  & 5.0\%  & [2.1\%,~~8.9\%]  & 95.0\% \\
10\% & 10.0\% & [5.7\%, 15.0\%]  & 90.0\% \\
20\% & 20.0\% & [14.0\%, 26.6\%] & 80.0\% \\
30\% & 30.0\% & [23.0\%, 37.3\%] & 70.0\% \\
\bottomrule
\end{tabular}
\end{table}

\noindent\textbf{Finding 4.2: \textsc{Centroid Pooled}
does not favor cheap or expensive instances.}
Our selection optimizes for representativeness
rather than cost, it neither favors cheap instances
nor avoids expensive ones. Table~\ref{tab:cost_savings} reports random draws. \textsc{Centroid Pooled} is deterministic, so it produces one subset per temporal split rather than a distribution of draws, and we report it separately. Its 10\% subset consumes 10.48\% of the run's token cost, averaged over all temporal splits, which sits inside the middle 95\% of the random draws at the same subset size and shows no detectable cost bias. In other words, the accuracy and risk improvements from
Sections~\ref{sec:results_rq1}--\ref{sec:results_rq3}
come with proportional cost savings. For example, a
10\% \textsc{Centroid Pooled} subset reduces MaxErr by 11\% against the typical baseline draw and cuts token consumption from 3.44B to
roughly 345M.

\medskip
\noindent\fbox{%
\parbox{\dimexpr\linewidth-2\fboxsep-2\fboxrule}{%
\textbf{Summary of RQ4:} Subset cost is proportional to subset size, i.e., a $k$\% subset consumes $k$\% of the
full benchmark cost on average across 10{,}000 random
draws per run, despite per-instance cost varying up
to 492$\times$. \textsc{Centroid Pooled} shows no detectable cost bias: its 10\% subset consumes 10.48\% of the full benchmark's token cost, within the range of the random draws. The RMSE and MaxErr gains come with
proportional cost savings. A 10\% subset reduces
token consumption from 3.44B to roughly 345M.
}%
}

%% file: Sections/Implications.tex
\section{Implications}
\label{sec:discussion}

Selecting subsets of large test suites is important
when a full evaluation is too expensive to run on every change. Developers run the full test suite on a fixed cadence, for example once a week, and need a cheaper check for the changes they make in between. After a model, prompt, or agent change, our
proposed approach can select a small subset (e.g.,
10\%) that provides a faster regression signal while
preserving the full test suite's historical difficulty and
typical agent trajectory profile.

\subsection{Implications for Developers}

\subsubsection{When Embeddings Matter Most}\label{sec:when_embeddings_matter}
Practitioners face a recurring choice: how small can
the evaluation subset be before the savings stop
being worth the loss in reliability? Embeddings matter most when the subset is small. The gap between trajectory-aware and baseline selection shrinks as the subset grows. Below 20\%, baseline selection produces larger errors in both the average case and the worst case, so embedding-based selection is the safer option.
At 30\%, \textsc{Consist-Strat} 
matches the average estimation error of
\textsc{Centroid Pooled}, so the embedding pipeline is justified when both lower RMSE and lower MaxErr are needed, for example
when a wrong resolve-rate estimate could cause a team to revert a good agent update or miss a real performance drop.

\subsubsection{Accuracy and Risk Are Controlled Separately}
A subset selection method has two failure modes: it can be wrong on average (high RMSE), or it can produce occasional errors that are much larger than typical (high MaxErr). RQ1 shows that outcome stratification alone (\textsc{Consist-Strat}) reduces RMSE but leaves MaxErr exposed to sampling variance. RQ3 shows that replacing the random draw with a fixed selection is not enough on its own: a fixed outcome-based selection without embeddings (\textit{Stability-Stratified}) helps only at the smallest subset sizes and falls behind random sampling from 20\% up. \textsc{Centroid Pooled} lowers both RMSE and MaxErr at 5\% and 10\%, so a team that needs either faithful average estimates or protection against large errors needs the embedding-based selection within outcome groups rather than a cheaper deterministic rule.

\subsubsection{Operational Guidelines}

\paragraph{Picking a subset size.}
Subset size is a tunable parameter, and Section~\ref{sec:when_embeddings_matter} gives the accuracy trade-off. The practical consequence is a fallback: at 20--30\%, a team without a trajectory pipeline can use \textsc{Consist-Strat} and lose little in RMSE, accepting the larger occasional errors that \textsc{Centroid Pooled} avoids. Below 20\%, there is no equivalent fallback.

\paragraph{Changing the agent architecture.}
When a team changes the agent architecture (the scenario evaluated in the \textbf{Multi-agent} dataset), the agent's behavior
on the test suite can shift in ways the historical
trajectories do not capture. A change of this size is one a team would typically follow with a full run, both to get a trustworthy number and to refresh the trajectories the selection depends on. Subset selection still applies to the iterations that come before that full run. \textsc{Centroid
Pooled} picks the most typical instances, so it may
miss the unusual failure modes that a new agent
introduces. \textsc{Core/Edge Pooled} keeps most of
its picks typical but reserves a small share for the
most unusual instances in the embedding space, making it a better fit when the agent architecture changes and unusual behaviors are expected.

\paragraph{When to refresh the embeddings.}
The embeddings capture what it takes for the agent to solve typical instances, so the selected subset represents the test suite. If a change to the agent
or the test suite is large enough to change how the
agent solves typical instances, the embeddings no
longer reflect current behavior and need to be
refreshed by re-running the embedding pipeline on the
full test suite.

\paragraph{How much history is needed.}
Our approach needs at least one past run, more runs improve both average and worst-case error. With one past run, instances can
only be grouped into pass or fail. The grouping is
too coarse to lower average error, but it still
reduces worst-case error, so a team that mainly wants
to avoid large mistakes can deploy our approach with a single past run. With two or more past runs, instances can
be grouped by how often they passed, which produces
stable improvements in both average and worst-case
error across different agents. Three or more runs is
the recommended target. However, runs that are too old may no longer reflect how the agent behaves. If full runs happen infrequently, using only the two or three most recent runs avoids relying on outdated trajectories.

A team that runs evaluations regularly accumulates
history over time, dropping outdated runs and refreshing the strata and embeddings with recent ones. Consider an organization
shipping an agent with a weekly release
cycle. Before each weekly release, the team runs the
full test suite to validate the new version. The full
run produces a fresh set of trajectories and
outcomes, which feed back into the history. During the week,
developers iterate on prompts, models, or
architectures, and use the selected subset as a fast
regression check in their pipeline, so every
pull request can be tested in a fraction of the time
and cost of the full test suite before being merged.
The next weekly release run then refreshes the
history again, and the cycle continues.

\subsection{Implications for Researchers}
Our findings open several research directions:
\begin{itemize}
    \item The sanitization pipeline requires per-dataset calibration (Sections~\ref{subsubsec:sanitization1} and~\ref{subsubsec:sanitization2}). Automating the detection and removal of outcome leakage across new benchmarks would make the approach easier to adopt.
    \item Pooled embeddings matches or outperforms time-series embeddings in most configurations (Table~\ref{tab:rq2-embedding}). Better temporal representations may produce different results.
    \item Our evaluation covers SWE-Bench Verified and SWE-Rebench. Applying the approach to other agent benchmarks, such as web navigation or tool-use benchmarks, would test whether the findings generalize beyond code-editing instances.
    \item The selected subset is fixed between full runs. Incremental updating, where the subset adapts as new partial results arrive, could reduce the need for periodic full re-runs.
\end{itemize}

%% file: Sections/ThreatsToValidity.tex
\section{Threats to Validity}
\label{sec:threatstovalidity}

\subsection{External Validity}
Our evaluation could overfit to a narrow setting,
such as a specific agent, model, or test suite
difficulty level. To reduce the overfitting risk, we
evaluate across three datasets that cover different
sources of variation: 45 reruns of the same agent
(\textbf{Single-setup}), 6 runs that change the model
or configuration (\textbf{Multi-model}), and 7 runs
that change the agent framework
(\textbf{Multi-agent}). For the \textbf{Multi-model} and \textbf{Multi-agent} datasets, we generate 1,000 synthetic distributions spread across nine source run difficulty levels (10\% to 90\%, approximately 111 per level, as described in Section~\ref{subsec:synthetic_distribution}), so the results do not depend on a single difficulty level. Both benchmarks are code-repair benchmarks built from GitHub issues, so the findings may not transfer to agent benchmarks in other domains.

\subsection{Construct Validity}
We measure subset quality using RMSE and MaxErr of
the estimated resolve rate, which captures how well a
subset tracks the full-suite resolve rate but does
not directly measure whether the subset reflects what
the full test suite requires from the agent. To
address the gap, we evaluate the estimate under conditions where a subset that is representative only by coincidence would fail. Every subset is selected from past runs and evaluated on later runs it has never seen (Section~\ref{sec:eval_setup}), across three scenarios that change the model, the execution settings, and the agent framework, so the subset has to stay representative as the agent's behavior changes. We also report two metrics that capture different failure modes, so a subset that tracks the resolve rate on average but omits behavioral profiles is still penalized through MaxErr. RQ3 confirms that the estimate depends on the subset preserving the full test suite's difficulty composition: removing outcome grouping raises RMSE by 30--37\%, and replacing it with behavioral clustering raises RMSE by 9--16\%.

\subsection{Internal Validity}

\paragraph{Look-ahead bias in subset selection.}
A subset selection method that uses information from
future runs would produce optimistic results. To
prevent look-ahead bias, all evaluations use a
temporal split: for a given window size $W$, $W$ past
runs are used to select the subset, and the subset is
evaluated on runs that happen strictly after the
latest past run in the window. The past runs provide the pass/fail outcomes and the trajectories used for selection, so no future-run outcome is visible at selection time. For the \textbf{Multi-model} and \textbf{Multi-agent} datasets,
we order runs by their submission date to the SWE-bench
experiments repository. Each submission represents a developer
updating the agent, so submission order reflects a realistic
development timeline. For the \textbf{Single-setup} dataset, no
natural chronological order exists among reruns, so any
permutation is equally valid.

\paragraph{Difficulty levels estimated from one run.}
Estimating the difficulty levels from a single source run (Section~\ref{subsec:synthetic_distribution}) could misrank distributions on other runs. We verified that it does not: on all remaining runs of both the \textbf{Multi-model} and \textbf{Multi-agent} datasets, the ordering of difficulty levels never inverts. Spearman's rank correlation between the source-run difficulty level and each remaining run's resolve rate is $\rho \geq 0.99$. On every run, distributions within the same difficulty level land within about 2 percentage points of each other, while neighboring levels sit 5--6 percentage points apart, so the levels do not overlap. Individual instances do change outcome between runs. Comparing any two runs of the \textbf{Multi-model} dataset, 21\% of instances resolve on one run and fail on the other, on average across all run pairs. The change is mostly in one direction, since the newer models resolve instances the older ones failed, and it concentrates in the lowest difficulty levels. Because our metrics compare subset and population resolve rates within the same run, these changes do not bias the estimation.

\paragraph{Outcome leakage in the embeddings.}
If the trajectory text contains words that directly
reveal whether a run passed or failed, the embeddings
would group instances by outcome rather than by agent
behavior, which would bias the selection. To prevent
outcome leakage, we sanitize the trajectory text in
four phases before embedding
(Section~\ref{subsec:sanitization}) and confirm with a
TF-IDF probe Section~\ref{subsubsec:sanitization4}) that the sanitized text no longer contains explicit outcome words. The probe still predicts pass/fail at 66--80\% accuracy from task-related words, so the sanitization removes the explicit indicators, not every correlation with outcome.

\paragraph{Instability of stochastic baselines.}
Each stochastic baseline produces a different subset on every run. With too few runs, the observed performance may reflect a lucky or unlucky set of draws rather than the method's actual quality. To stabilize the baseline
estimates, we run each stochastic method with 500
independent seeds for every distribution and temporal split, and confirm at 500, 1{,}000, and 2{,}000 seeds that the rankings and effect sizes do not change. Each seed draws a fresh stratified random sample with an independent \texttt{numpy} random generator, so no seeds overlap across methods or temporal splits. To keep the MaxErr comparison fair between stochastic and deterministic methods, we report per-seed MaxErr statistics (median, P95, worst) rather than a single maximum over all 500 seeds.

\subsection{Conclusion Validity}
\label{subsec:conclusion_validity}
Conclusion validity concerns whether the statistical conclusions drawn from the data are correct~\cite{cook1979quasi}. Such conclusions can be wrong:
$p$-values can be inflated by running many
comparisons, the chosen test may not fit the data,
and dependence between samples can make $p$-values optimistic.

To control for inflated $p$-values from running many
comparisons, we apply Holm-Bonferroni
correction~\cite{holm1979simple} across the full
comparison family. A low $p$-value shows that a difference exists but not whether it is large enough to matter in practice. We report Cliff's $\delta$~\cite{romano2006appropriate} alongside $p$-values to measure the practical size of each difference, following Benavoli et al.~\cite{benavoli2017time}.

To avoid distributional assumptions that may not hold
for our error metrics, all pairwise comparisons use
the two-sided Wilcoxon signed-rank
test~\cite{wilcoxon1945individual}, which ranks
differences without assuming normality.

The 1,000 synthetic distributions are drawn from a pool of 500 instances, so any two distributions share about a third of their instances (Section~\ref{subsec:synthetic_distribution}). The Wilcoxon test treats the distributions as independent pairs, which makes its $p$-values optimistic. No subset of independent distributions exists in this design: greedy selection of distributions with pairwise Jaccard below 0.30 retains only 3 of the 1,000 on the \textbf{Multi-model} dataset, because two 250-instance subsets of a 500-instance pool share 125 instances on average. We address the threat in two ways. First, our conclusions rest on Cliff's $\delta$, whose point estimate counts paired wins and is not inflated by dependence, and we treat a negligible $\delta$ as no difference. Second, we repeat the RQ2 comparison at the level of the nine difficulty levels, treating each level as one unit and asking whether the median paired RMSE difference within the level favors the trajectory-aware method. On the \textbf{Multi-model} dataset, \textsc{Core/Edge Pooled} beats \textsc{Consist-Strat} on all nine levels at both 5\% and 10\% subset size (sign test, $p = 0.004$), and \textsc{Centroid Pooled} on all nine at 5\% and eight of nine at 10\% ($p = 0.004$ and $p = 0.039$). The Table~\ref{tab:rq2-embedding} ranking therefore holds within nearly every difficulty level and is not driven by a few highly overlapping distributions.

%% file: Sections/Conclusion.tex
\section{Conclusion}
\label{sec:conclusion}

Developers of SWE-agents rerun a full benchmark after every change to the model, prompt, or architecture, which is too expensive to do on each change. In this study, we introduce a trajectory-aware
subset selection approach for SWE-agent regression
testing that combines historical outcome
stratification with deterministic geometric selection
over sanitized trajectory embeddings. Through four research questions, we evaluate 76 subset selection configurations, spanning baselines, trajectory-aware selection, and ablations, across three regression scenarios on SWE-Bench Verified and SWE-Rebench.
At a 10\% subset size, our approach achieves a median estimation error below 5\% and a worst-case error under 10\%, reducing worst-case error by 4–11\% against the typical baseline draw while maintaining lower average error by up to 11\%. The improvement comes with
a 90\% reduction in token cost, making the approach a practical replacement for full-suite evaluation between the periodic full runs of a SWE-agent under active development. Ablation analysis confirms that outcome stratification and geometric selection are both necessary. Outcome stratification controls the difficulty composition of
the subset, and deterministic geometric selection
controls behavioral representativeness within each
outcome group. Future research can build on our study by
studying when the trajectory embeddings should be recomputed as the agent and the test suite change, and by extending the
approach to agent domains beyond software
engineering.